\documentclass[aip,reprint,floatfix]{revtex4-1}
\usepackage{graphicx}
\usepackage{dcolumn}
\usepackage{bm}

\usepackage[utf8]{inputenc}
\usepackage[T1]{fontenc}
\usepackage{mathptmx}
\usepackage{multirow}
\usepackage{booktabs}
\draft 

\begin{document}


\title{Resonant effect at the ultrarelativistic 
electron-positron pairs production by gamma quanta in the field of a nucleus and a pulsed light wave}

\author{Sergei P. Roshchupkin}
\email[]{serg9rsp@gmail.com}
\affiliation{Department of Theoretical Physics, Peter the Great St.Petersburg Polytechnic University, Saint-Petersburg, Russia}

\author{Nikita R. Larin}
\email[]{nikita.larin.spb@gmail.com}
\affiliation{Department of Theoretical Physics, Peter the Great St.Petersburg Polytechnic University, Saint-Petersburg, Russia}

\author{Victor V. Dubov}
\email[]{maildu@mail.ru}
\affiliation{Department of Theoretical Physics, Peter the Great St.Petersburg Polytechnic University, Saint-Petersburg, Russia}


\date{\today}

\begin{abstract}
Resonant electron-positron pair production by a high-energy gamma quantum ${\omega _i} \mathbin{\lower.3ex\hbox{$\buildrel>\over
{\smash{\scriptstyle\sim}\vphantom{_x}}$}} {10^2}\:{\rm{ GeV}}$ in the field of a nucleus and a quasi-monochromatic laser wave with the intensities $I \mathbin{\lower.3ex\hbox{$\buildrel<\over
{\smash{\scriptstyle\sim}\vphantom{_x}}$}} {10^{16}} \div {10^{17}}{{\rm{W}} \mathord{\left/
 {\vphantom {{\rm{W}} {{\rm{c}}{{\rm{m}}^{\rm{2}}}}}} \right.
 \kern-\nulldelimiterspace} {{\rm{c}}{{\rm{m}}^{\rm{2}}}}}$ was theoretically studied. Under the resonant condition an intermediate virtual electron (positron) in the laser field becomes a real particle. Due to that fact the initial process of the second order in the fine structure constant in a laser field effectively reduces into two successive processes of the first order: the laser-stimulated Breit-Wheeler process and the laser-assisted process of an intermediate electron (positron) scattering by a nucleus. It is shown that there is a threshold energy for the initial gamma quantum, which significantly depends on the number of absorbed photons of a wave. In the resonant condition the electron-positron pair energy is determined by the outgoing angle of a positron (for the channel A) or an electron (for the channel B) relative to the initial gamma quantum momentum. The differential cross sections for the first few resonances with simultaneous registration of the energy and the outgoing angle of a positron or an electron were obtained. For the initial gamma quantum energy  ${\omega _i} = 125\;{\rm{GeV}}$ the resonant energies of an electron-positron pair for the case of first three resonances can be measured with a very high magnitude of the differential cross section: from $ \sim {10^{13}}$ for the first resonance to $ \sim {10^8}$  (in the units of $\alpha {Z^2}r_e^2$) for the third resonance. 
\end{abstract}


\maketitle

\section{Introduction}
Nowadays the plenty of powerful laser radiation sources are widely used in the physical experiment practice \cite{mourou,Phelix,Bula,Burke,Kanya}. Therefore the theoretical study of the quantum electrodynamics (QED) processes in a strong light field is one of the great priority trends which intensively develops \cite{Ritus,Fedorov,monmon,monpul,PiazzaRev,Ehlotzky_2009,Ehlotzky_rev,Nonres,intech,Larin,Kra,Oleinik1,Oleinik2,Dubov_2020,PPPNonRes,MultiPPP,PPPInter,Muller,AugustinBH,ChannelingBH,Kuchiev2007,Milstein2006, PiazzaStrong,Bunkin_Fedorov,Lebed__2016,Nar_fof,Res_rev_rsp,Zhou,krainov_rsp_slow,lebedev72,borisovSBRES,rspNucFiz,rsp_2002,Dondera,Florescu1,zhelt,zhelt2,Li,ErikLotsted,SchnezBrem,Nonres_sb,Leb_sb_NN,Rsp_lys_two,Rsp_lys_two2,pie_sb,Lebed2018,Krachkov_2019,PhysRevA.81.033413}. The main results were systematized in monographs \cite{Ritus,Fedorov,monmon,monpul} and reviews \cite{PiazzaRev,Ehlotzky_2009,Ehlotzky_rev,Nonres,intech}. 

It is important to emphasize that QED processes of higher than the first order in the fine structure constant in a laser field (the QED processes assisted by a laser field) can occur through the resonant channels. In a laser field so-called Oleinik resonances \cite{Oleinik1,Oleinik2} can take place due to the fact that in an external light field the  first order processes in the fine structure constant (the laser-stimulated processes of QED) are allowed \cite{Ritus}. We accentuate that probability of the resonant process in a laser field significantly (by the several orders of magnitude) exceeds the corresponding one for the process without a laser field. 

 There are many previous researches \cite{Larin,intech,monmon,monpul,Kra} dedicated to the problem of resonant photoproduction of pair (PPP). For example, there was considered the resonance only for one of the possible channels, when in the field of a wave the initial gamma quantum products a positron and an intermediate electron, which is then scattered by a nucleus (the channel A) \cite{monmon,monpul,intech}. The second channel, when the initial gamma quantum products an electron and an intermediate positron, which is then scattered by a nucleus (the channel B), has not been studied. Also, we note that for channel A there was examined the case of an ultrarelativistic pair, where positron propagates in a narrow cone with the direction of the initial gamma quantum momentum but an electron is scattered at a large angle. In the article \cite{Larin} authors investigated first resonance (with the absorption of one photon of a wave) of the PPP with the ultrarelativistic energies of a pair, when a positron and an electron propagate in a narrow cone along the initial gamma quantum momentum direction with taking into account the channels A and B. This process was studied in the plane monochromatic wave field. It should be noted that the crossed channel of the concerned process is a spontaneous bremsstrahlung of an ultrarelativistic electron in the field of a nucleus and a plane monochromatic wave \cite{Dubov_2020}. In addition, we would like to point out some works where authors draw attention to the nonresonant PPP and discuss related issues
\cite{PPPNonRes,MultiPPP,PPPInter,Muller,AugustinBH,ChannelingBH,Kuchiev2007,Milstein2006, PiazzaStrong}.

In the present paper, we develop the theory of several first resonances (with the absorption of one, two, three and so on photons of a wave) for the process of an electron-positron pair production by a high-energy gamma quantum in the field of a nucleus and a pulsed laser wave. 

There are two characteristic parameters in the problem of PPP on a nucleus in the field of a plane wave. Classical relativistically invariant parameter is \cite{Ritus}

\begin{eqnarray}
{\eta _0} = \frac{{e{F_0}\mathchar'26\mkern-10mu\lambda  }}{{m{c^2}}},
\label{1}
\end{eqnarray}
which numerically equals to the ratio of the work of a field at a wavelength to the electron rest energy ($e$ and $m$ are the charge and the electron mass, ${F_0}$ and $\:\mathchar'26\mkern-10mu\lambda  = {c \mathord{\left/
 {\vphantom {c \omega }} \right.
 \kern-\nulldelimiterspace} \omega }$ are the strength and electromagnetic wavelength, $\omega $ is the wave frequency). The quantum multiphoton parameter \cite{Bunkin_Fedorov,Fedorov} (Bunkin-Fedorov parameter) is
 
 \begin{eqnarray}
{\gamma _0} = {\eta _0}\frac{{mvc}}{{\hbar \omega }}.
\label{2}
\end{eqnarray}
Herein ${v}$ is the electron (positron) velocity, ${c}$ is a light speed. Within the optical frequencies range ({$\omega \sim{10^{15}}{s^{ - 1}}$}) the classical parameter is of the order of unity ${\eta _0}\sim1$ for the fields ${F_0}\sim{10^{10}} \div {10^{11}}\;{V \mathord{\left/
 {\vphantom {\rm{V} {cm}}} \right.
 \kern-\nulldelimiterspace} \rm{{cm}}}$. At the same time the quantum parameter has the order of ${\gamma _0}\sim1$ for the fields ${F_0}\sim\left( {{{10}^5} \div {{10}^6}} \right)\left( {{c \mathord{\left/
 {\vphantom {c v}} \right.
 \kern-\nulldelimiterspace} v}} \right)\;{{\rm{V}} \mathord{\left/
 {\vphantom {{\rm{V}} {{\rm{cm}}}}} \right.
 \kern-\nulldelimiterspace} {{\rm{cm}}}}$. Hence, in the fields with ${\eta _0}\sim1$ the quantum parameter ${\gamma _0}$  may be large. However, it is true only if electrons (positrons) are scattered by a nucleus at large angles. In such situation the quantum parameter ${\gamma _0}$ defines multiphoton processes \cite{Bunkin_Fedorov,Fedorov}. We underline that for the process of PPP when the electrons (positrons) are scattered by a nucleus at small angles, the quantum parameter (\ref{2}) does not appear \cite{Lebed__2016,Larin}. Consequently, the main parameter which determines the multiphoton processes is the classical relativistically invariant parameter (\ref{1}). This case will be considered in the present article. Throughout this paper we will use the relativistic system of units: $\hbar  = c = 1$.
 
 \section{The Amplitude of the Process}
 Let us choose the four-potential of a plane electromagnetic pulse wave propagating along the $z$ axis in the following form:
 \begin{eqnarray}
A\left( \varphi  \right) = \left( {\frac{{{F_0}}}{\omega }} \right) \cdot g\left( {\frac{\varphi }{{\omega \tau }}} \right) \cdot \left( {{e_x}\cos\varphi  + \delta {e_y}\sin \varphi } \right), \nonumber\\
\varphi  = kx = \omega \left( {t - z} \right),
\label{3}
\end{eqnarray}
where $k = \left( {\omega ,{\bf{k}}} \right)$ is the wave four-vector, $\delta $ is the ellipticity parameter of a wave. Also here ${e_x} = \left( {0,{{\bf{e}}_x}} \right)$, ${e_y} = \left( {0,{{\bf{e}}_y}} \right)$ are the polarization four-vectors of the wave, particularly $e_{x,y}^2 =  - 1,{\rm{ }}\left( {{e_{x,y}}k} \right) = {k^2} = 0$. In the expression (\ref{3}) the function $g\left( {{\varphi  \mathord{\left/
 {\vphantom {\varphi  {\omega \tau }}} \right.
 \kern-\nulldelimiterspace} {\omega \tau }}} \right)$ is the envelope of our potential. We require $g\left( 0 \right) = 1$ and the exponential decreasing of the function $g \to 0$ when $\left| \varphi  \right| \gg \omega \tau $. Hereinafter, we assume that the duration of a pulse significantly exceeds the characteristic oscillation time of a wave 
  \begin{eqnarray}
\omega \tau  \gg 1.
\label{4}
\end{eqnarray}
In this case we are able to consider  ${\tau}$ as a duration of a laser pulse. We treat the effect of the nuclear Coulomb potential in the Born approximation. This consideration leads to the restrictions on nuclear charge (${{Z{e^2}} \mathord{\left/
 {\vphantom {{Z{e^2}} {v <  < 1}}} \right.
 \kern-\nulldelimiterspace} {v \ll 1}}$, $Z$ is the nuclear charge).
 
\begin{figure}[ht]
\includegraphics[width=1\linewidth]{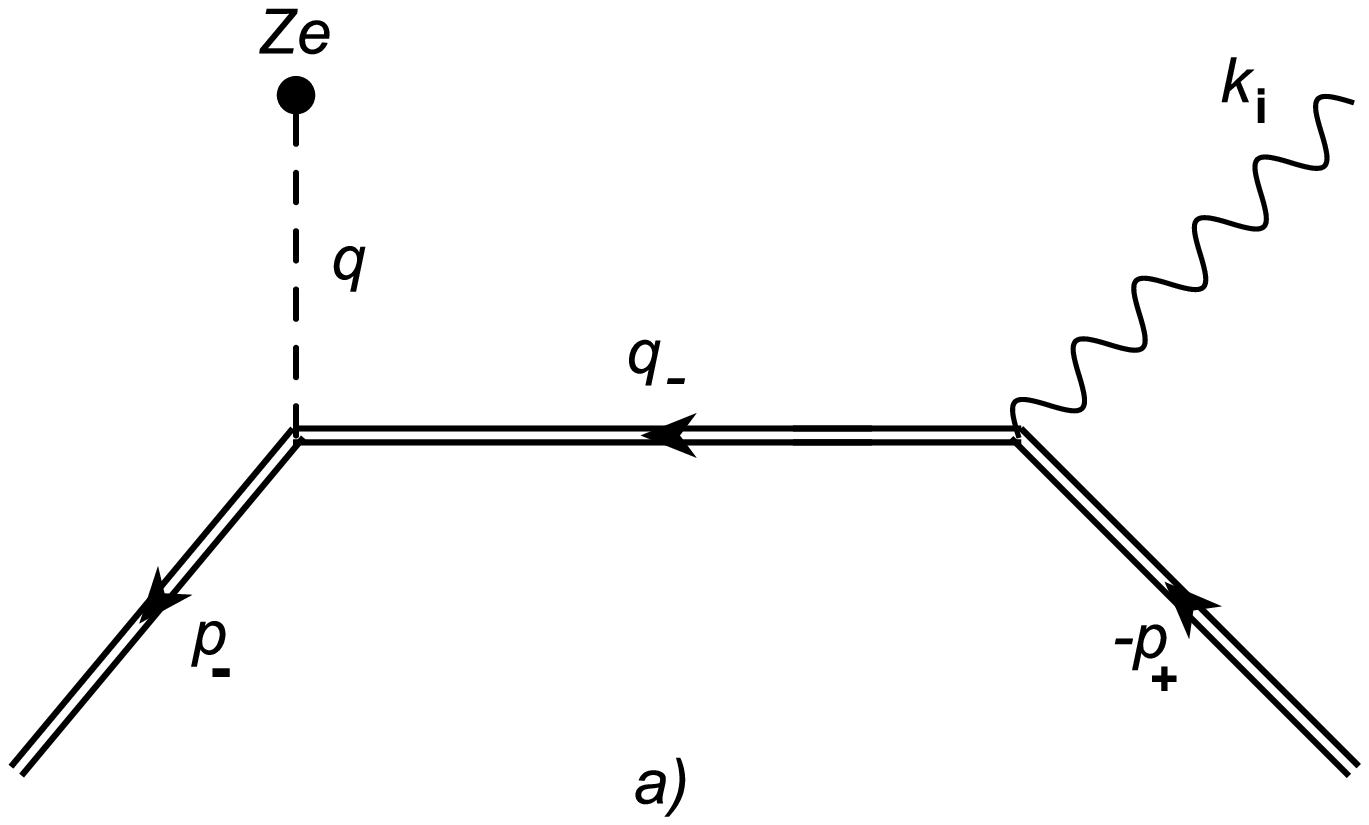}%
\vfill
\includegraphics[width=1\linewidth]{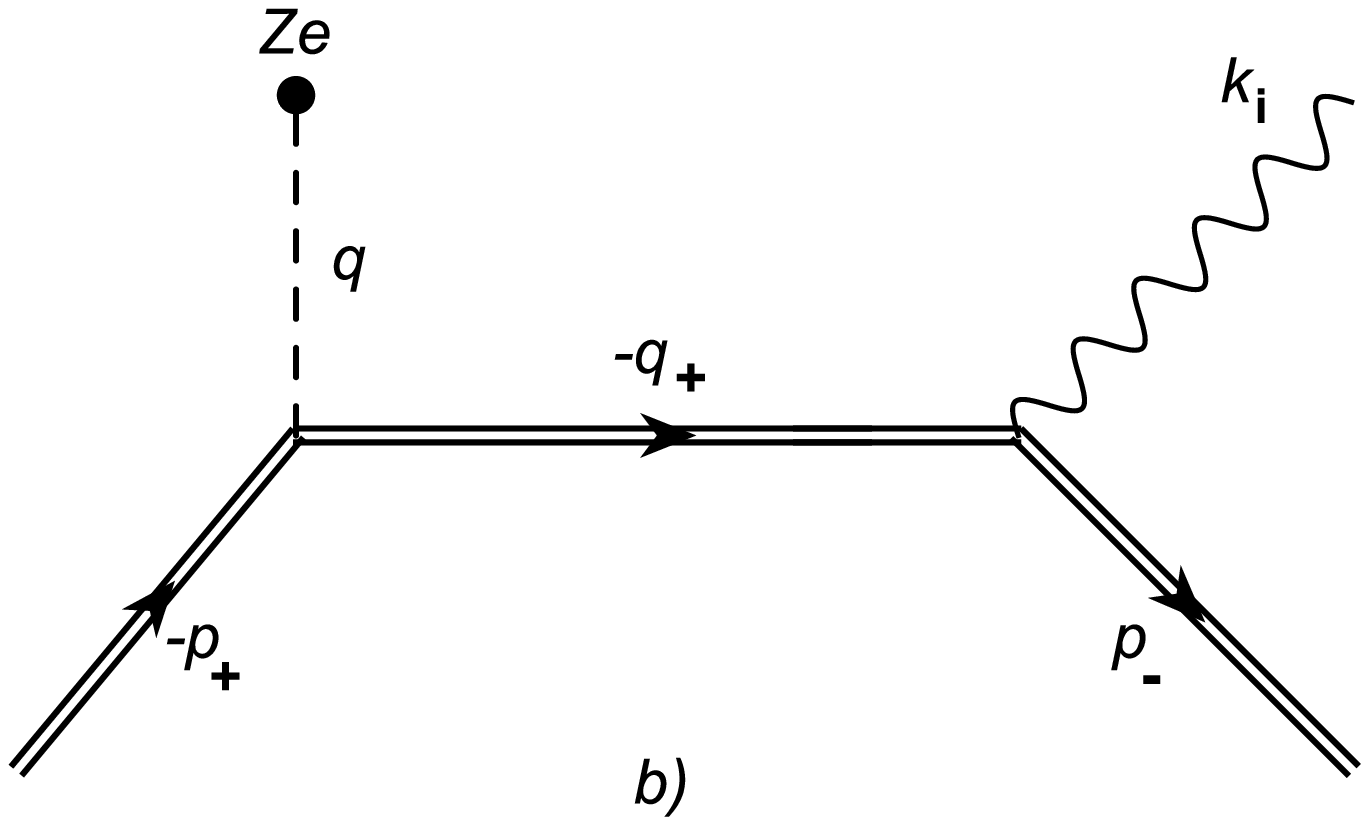}%
\caption{ Feynman diagrams of the PPP process on a nucleus in the field of a pulsed light wave. Double incoming and outgoing lines correspond to the Volkov functions of an electron and a positron in the initial and final states, the inner lines correspond to the Green function of the fermion in the field of a plane wave (\ref{3}). Wavy lines correspond to the initial gamma quantum. Dashed lines stand for the pseudo photon of recoil.\label{Fig1}}%
\end{figure}

 The process of PPP on a nucleus in the presence of a light field is the second order process in the fine structure constant and it is described by two Feynman diagrams (see Fig.\ref{Fig1}).
 
The envelope function is selected as a function of variable $\varphi $, therefore the electromagnetic field (\ref{3}) can be treated as a plane wave. As known, there is an exact solution of the Dirac equation for an electron in the field of a plane wave with the arbitrary spectral composition (Volkov function) \cite{Volkov}. Also there is the expression for the Green's function of an electron in the field of a plane wave \cite{BrownKibble,Schwinger}.

For further calculations, we consider the process in the case of a circular polarization $\left( {{\delta ^2} = 1} \right)$ and the field intensity in the pulse peak is
\begin{eqnarray}
{\eta _0} \ll 1.
\label{5}
\end{eqnarray}

Taking this consideration into account \cite{intech,monpul}, we obtain the expression for the amplitude of the PPP process on a nucleus in the field of a plane quasi-monochromatic (\ref{4}) weak electromagnetic wave (\ref{5}):
\begin{eqnarray}
S = \sum\limits_{l =  - \infty }^{ + \infty } {{S_{(l)}}}.
\label{6}
\end{eqnarray}
Herein ${S_{(l)}}$ is a partial amplitude of the process with radiation (absorption) of $\left| l \right|$-photons of a wave
\begin{eqnarray}
{S_{\left( l \right)}} = i\frac{{2{\tau ^2}\omega Z{e^3}\sqrt \pi  }}{{\sqrt {2{\omega _i}{E_ - }{E_ + }} }}\frac{{\left( {{{\bar u}_{{p_ - }}}{{B'}_{\left( l \right)}}{v_{ - {p_ + }}}} \right)}}{{\left[ {{{\bf{q}}^2} + {q_0}\left( {{q_0} - 2{q_z}} \right)} \right]}},
\label{7}
\end{eqnarray}
where
\begin{eqnarray}
{B'_{(l)}} = {B'_{ + \left( l \right)}} + {B'_{ - \left( l \right)}},\quad q = {p_ - } + {p_ + } - {k_i} + lk.
\label{8}
\end{eqnarray}
Here $q = \left( {{q_0},{\bf{q}}} \right)$ is a transferred four-momentum, ${B_{ + \left( l \right)}}$ and ${B_{ - \left( l \right)}}$ are the amplitudes for the channels A and B, ${v_{ - {p_ + }}}$ and ${\bar u_{{p_ - }}}$ are the Dirac bispinors for a positron and an electron,
\begin{eqnarray}
{B_{ + \left( l \right)}} = \sum\limits_{r =  - \infty }^{ + \infty } {\int\limits_{ - \infty }^\infty  {d{\phi _1}} \int\limits_{ - \infty }^\infty  {d{\phi _2}\exp \left( {i{q_0}\tau {\phi _2}} \right)} M_{(l + r)}^0\left( {{p_ - },{q_ - },{\phi _2}} \right)}\times\nonumber\\
\times G\left( {{q_ - },{\phi _1} - {\phi _2}} \right)\left[ {\varepsilon _\mu ^{}F_{( - r)}^\mu \left( {{q_ - },{p_ + },{\phi _1}} \right)} \right],\quad\quad
\label{9}
\end{eqnarray}
\begin{eqnarray}
{B_{ - \left( l \right)}} = \sum\limits_{r =  - \infty }^{ + \infty } {\int\limits_{ - \infty }^\infty  {d{\phi _1}} \int\limits_{ - \infty }^\infty  {d{\phi _2}\exp \left( {i{q_0}\tau {\phi _2}} \right)} M_{(l + r)}^0\left( {{p_ + },{q_ + },{\phi _2}} \right)}\times\nonumber\\
\times G\left( {{q_ + },{\phi _1} - {\phi _2}} \right)\left[ {\varepsilon _\mu ^{}F_{( - r)}^\mu \left( {{q_ + },{p_ - },{\phi _1}} \right)} \right],\quad\quad
\label{10}
\end{eqnarray}
\begin{eqnarray}
G\left( {{q_ \pm },{\phi _1} - {\phi _2}} \right) =\int\limits_{ - \infty }^\infty  {d\xi } \left[ {\frac{{\left( {{{\hat q}_ \pm } + m} \right) + \xi \hat k}}{{\left( {q_ \pm ^2 - {m^2}} \right) + 2\xi \left( {k{q_ \pm }} \right)}}} \right]\times\nonumber\\
\times\exp \left[ {i\left( {\omega \tau \xi } \right)\left( {{\phi _1} - {\phi _2}} \right)} \right],
\label{11}
\end{eqnarray}
\begin{eqnarray}
{q_ - } =  - {p_ + } + {k_i} + rk,\quad {q_ + } =  - {p_ - } + {k_i} + rk.
\label{12}
\end{eqnarray}
The notations with hats stand for dot product of the corresponding four-vector with the Dirac gamma matrices: ${\tilde \gamma _\mu } = \left( {{{\tilde \gamma }_0},{\bf{\tilde \gamma }}} \right)$, $\mu  = 0,1,2,3$ (for example, ${\hat q_ \pm } = q_ \pm ^\mu {\tilde \gamma _\mu } = q_ \pm ^0{\tilde \gamma _0} - {{\bf{q}}_ \pm }{\bf{\tilde \gamma }}$). Besides ${\phi _n} = {{{\varphi _n}} \mathord{\left/
 {\vphantom {{{\varphi _n}} {\omega \tau }}} \right.
 \kern-\nulldelimiterspace} {\omega \tau }}{\rm{ }}\left( {n = 1,2} \right)$, ${\varepsilon ^\mu }$ and ${k_i} = \left( {{\omega _i},{{\bf{k}}_{\bf{i}}}} \right)$ are the polarization four-vector and the four-momentum of the initial gamma quantum correspondingly, ${p_ \pm } = \left( {{E_ \pm },{{\bf{p}}_ \pm }} \right)$ are the four-momenta of a positron and an electron, ${q_ - }$ and ${q_ + }$ are the four-momenta of an intermediate electron and positron for the channels A and B accordingly (see Fig.\ref{Fig1}). In the expressions (\ref{9}) and (\ref{10}) there is the matrix $M_{(l + r)}^0\left( {p',p,{\phi _2}} \right)$, which under the resonant conditions and with the expression in square brackets in the denominator (\ref{7}) taken into account represents the amplitude of the intermediate electron (positron) scattering on a nucleus $\left( {p \to p'} \right)$ with radiation (absorption) of $\left| {l + r} \right|$-photons of a wave (the laser-assisted Mott process) \cite{Bunkin_Fedorov,Fedorov}. At the same time $F_{( - r)}^\mu \left( {p',p,{\phi _1}} \right)$ represents the amplitude of an electron-positron pair production (with momenta $p$ and $p'$) by a gamma quantum ${k_i}$ with absorption of $r$-photons of a wave (the laser-stimulated Breit-Wheeler process) \cite{Ritus}
\begin{eqnarray}
M_{(l + r)}^0\left( {p',p,{\phi _2}} \right) = {\tilde \gamma ^0}{L_{l + r}}\left( {p',p,{\phi _2}} \right),
\label{13}
\end{eqnarray}
\begin{eqnarray}
F_{( - r)}^\mu \left( {p',p,{\phi _1}} \right) = {\tilde \gamma ^\mu }{L_{-r}}\left( {p',p,{\phi _1}} \right) + \nonumber\\
+b_{p'p( - )}^\mu \left( {{\phi _1}} \right){L_{-r-1}} + b_{p'p( + )}^\mu \left( {{\phi _1}} \right){L_{-r+1}}.
\label{14}
\end{eqnarray}
Here we introduced matrices $b_{p'p( \pm )}^\mu $ and special functions ${L_{n'}}\left( {p',p,{\phi _n}} \right)$ which were studied in details in the following article \cite{L_fun}:
\begin{eqnarray}
b_{p'p\left(  \pm  \right)}^\mu \left( {{\phi _1}} \right) = \eta \left( {{\phi _1}} \right)\left[ {\frac{m}{{4\left( {kp'} \right)}}{{\hat e}_ \pm }\hat k{{\tilde \gamma }^\mu } - \frac{m}{{4\left( {kp} \right)}}{{\tilde \gamma }^\mu }\hat k{{\hat e}_ \pm }} \right],\quad
\label{15}
\end{eqnarray}
\begin{eqnarray}
{L_s}\left( {p',p,{\phi _n}} \right) = \exp \left( { - is{\chi _{p'p}}} \right){J_s}\left[ {{\gamma _{p'p}}\left( {{\phi _n}} \right)} \right].
\label{16}
\end{eqnarray}
Herein ${J_s}$ is the Bessel function of an integer index. The parameters ${\gamma _{p'p}},{\rm{ }}{\chi _{p'p}}$ and four-vectors ${e_ \pm }$ are defined as follows:
\begin{eqnarray}
{\gamma _{p'p}}\left( {{\phi _n}} \right) = \eta \left( {{\phi _n}} \right)m\sqrt { - Q_{p'p}^2} ,{\rm{  }}\quad Q_{p'p}^{} = \frac{{p'}}{{\left( {kp'} \right)}} - \frac{p}{{\left( {kp} \right)}},\quad
\label{17}
\end{eqnarray}
\begin{eqnarray}
\tan {\chi _{p'p}} = \delta \frac{{\left( {{Q_{p'p}}{e_y}} \right)}}{{\left( {{Q_{p'p}}{e_x}} \right)}},{\rm{  }}{e_ \pm } = {e_x} \pm i\delta {e_y}.
\label{18}
\end{eqnarray}
In the relations (\ref{15})-(\ref{17}) the parameter $\eta \left( {{\phi _n}} \right)$ has the following form:
\begin{eqnarray}
\eta \left( {{\phi _n}} \right) = {\eta _0}g\left( {{\phi _n}} \right),\quad n = 1,2.
\label{19}
\end{eqnarray}
Here ${\eta _0}$  is the intensity of a wave in the pulse peak, $g\left( {{\phi _{1,2}}} \right)$ is the envelope function of a laser wave (\ref{3}). Herewith, the magnitudes of the four-momenta $p$ and $p'$ in the expressions (\ref{13})-(\ref{18}) are defined by the corresponding expressions in the amplitudes (\ref{9}) and (\ref{10}).

From the relation (\ref{11}) we can see that the substantive range of the variable $\xi $, which makes the main contribution to the integral (\ref{11}) is defined by the condition: 
\begin{eqnarray}
\left| \xi  \right| \mathbin{\lower.3ex\hbox{$\buildrel<\over
{\smash{\scriptstyle\sim}\vphantom{_x}}$}} \frac{1}{{\omega \tau }} \ll 1.
\label{20}
\end{eqnarray}
Since, when $\left| \xi  \right| \gg {1 \mathord{\left/
 {\vphantom {1 {\omega \tau }}} \right.
 \kern-\nulldelimiterspace} {\omega \tau }}$ due to the high-frequency oscillations, these integrals would be small. Thus, we can neglect $\xi \hat k$ in the denominator of the expression (\ref{11}) in comparison with $\left( {{{\hat q}_ \mp } + m} \right)$. It noteworthy that the dependence on the variable of integration in the denominator of the expression (\ref{11}) is a consequence of a pulsed behavior of the laser wave \cite{monpul,intech}. We emphasize that there is no such term in the case of a monochromatic wave and as a consequence that leads to the resonant infinity in the amplitude of the PPP process on a nucleus in the field of a wave\cite{monmon,Larin}. As a result, the integral (\ref{11}) is easily calculated and the expression for the channels A and B takes the following form:
 \begin{eqnarray}
G\left( {{q_ \pm },{\phi _1} - {\phi _2}} \right) = \frac{{\pi i\left( {{{\hat q}_ \pm } + m} \right)}}{{2\left( {k{q_ \pm }} \right)}}\exp \left[ { - 2i{\beta _ \pm }\left( {{\phi _1} - {\phi _2}} \right)} \right]\times\nonumber\\
\times{\mathop{\rm sgn}} \left( {{\phi _1} - {\phi _2}} \right).\quad
\label{21}
\end{eqnarray}
Herein ${\beta _ - }$ and ${\beta _ + }$ are the resonant parameters for the channels A and B \cite{monpul,intech}:
 \begin{eqnarray}
{\beta _ \mp } = \frac{{\left( {q_ \mp ^2 - m_{}^2} \right)}}{{4\left( {k{q_ \mp }} \right)}}\omega \tau.
\label{22}
\end{eqnarray}
Taking (\ref{21}) into account we finally deduce the expressions for the amplitudes (\ref{9}) and (\ref{10}):
\begin{eqnarray}
{B_{ + \left( l \right)}} = \sum\limits_{r =  - \infty }^\infty  {\frac{{\pi i}}{{\left( {k{q_ - }} \right)}}\int\limits_{ - \infty }^\infty  {d{\phi _1}\int\limits_{ - \infty }^\infty  {d{\phi _2}} \exp \left( { - 2i{\beta _ - }{\phi _1}} \right)} } \exp \left[ {i\left( {{q_0}\tau  + 2{\beta _ - }} \right){\phi _2}} \right]\times\nonumber\\
\times{\mathop{\rm sgn}} \left( {{\phi _1} - {\phi _2}} \right){M_{\left( {l + r} \right)}}\left( {{p_ - },{q_ - },{\phi _2}} \right)\left( {{{\hat q}_ - } + m} \right)\left[ {\varepsilon _\mu ^ * F_{( - r)}^\mu \left( {{q_ - },{p_ + },{\phi _1}} \right)} \right],\quad\quad
\label{23}
\end{eqnarray}
\begin{eqnarray}
{B_{ - \left( l \right)}} = \sum\limits_{r =  - \infty }^\infty  {\frac{{\pi i}}{{\left( {k{q_ + }} \right)}}\int\limits_{ - \infty }^\infty  {d{\phi _1}\int\limits_{ - \infty }^\infty  {d{\phi _2}} \exp \left( { - 2i{\beta _ + }{\phi _1}} \right)\exp \left[ {i\left( {{q_0}\tau  + 2{\beta _ + }} \right){\phi _2}} \right]} }\times\nonumber\\
\times{\mathop{\rm sgn}} \left( {{\phi _1} - {\phi _2}} \right){M_{\left( {l + r} \right)}}\left( {{p_ + },{q_ + },{\phi _2}} \right)\left( {{{\hat q}_ + } + m} \right)\left[ {\varepsilon _\mu ^ * F_{ - r}^\mu \left( {{q_ + },{p_ - },{\phi _1}} \right)} \right].\quad\quad
\label{24}
\end{eqnarray}
The expressions for the amplitudes (\ref{6})-(\ref{8}), (\ref{23}), (\ref{24}) are valid for the circular polarized quasi-monochromatic weak laser wave (\ref{4}), (\ref{5}).

From now on we will consider the high-energy initial gamma quanta and the ultrarelativistic energies of the produced electrons and positrons, when all particles propagate in a narrow cone along the direction of the initial gamma quantum momentum \cite{Larin}:
\begin{eqnarray}
{\omega _i} \gg m,\quad {E_ \pm } \gg m,
\label{25}
\end{eqnarray}
\begin{eqnarray}
{\theta _{i \pm }} = \angle \left( {{{\bf{k}}_i},{{\bf{p}}_ \pm }} \right) \ll 1,\quad {\bar \theta _ \pm } = \angle \left( {{{\bf{p}}_ + },{{\bf{p}}_ - }} \right) \ll 1,
\label{26}
\end{eqnarray}
\begin{eqnarray}
{\theta _i} = \angle \left( {{{\bf{k}}_i},{\bf{k}}} \right)\sim1,\quad {\theta _ \pm } = \angle \left( {{\bf{p}}_ \pm },{{\bf{k}}} \right)\sim1.
\label{27}
\end{eqnarray}

\section{POLES OF THE PPP AMPLITUDE}
\begin{figure}[ht]
\includegraphics[width=1\linewidth]{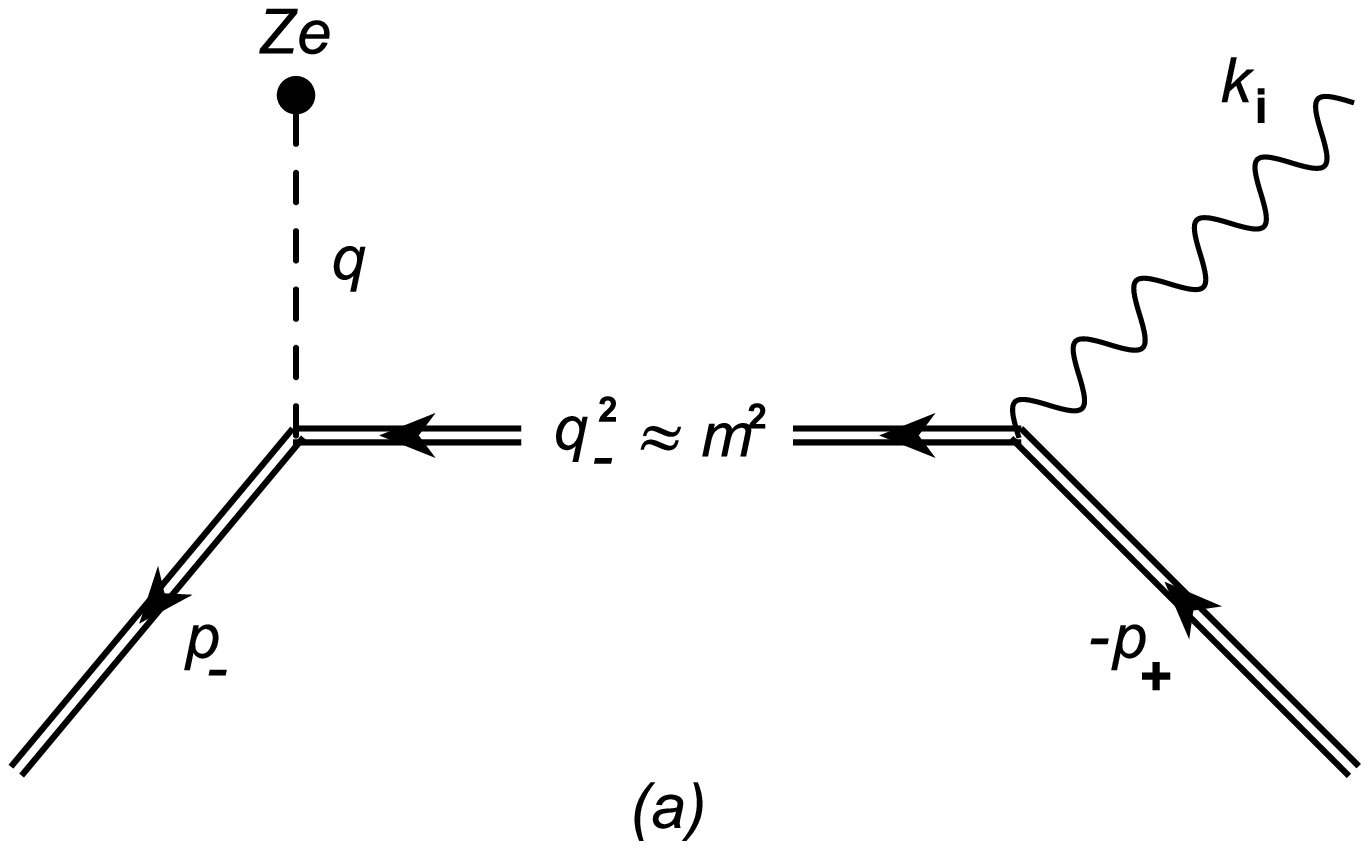}%
\vfill
\includegraphics[width=1\linewidth]{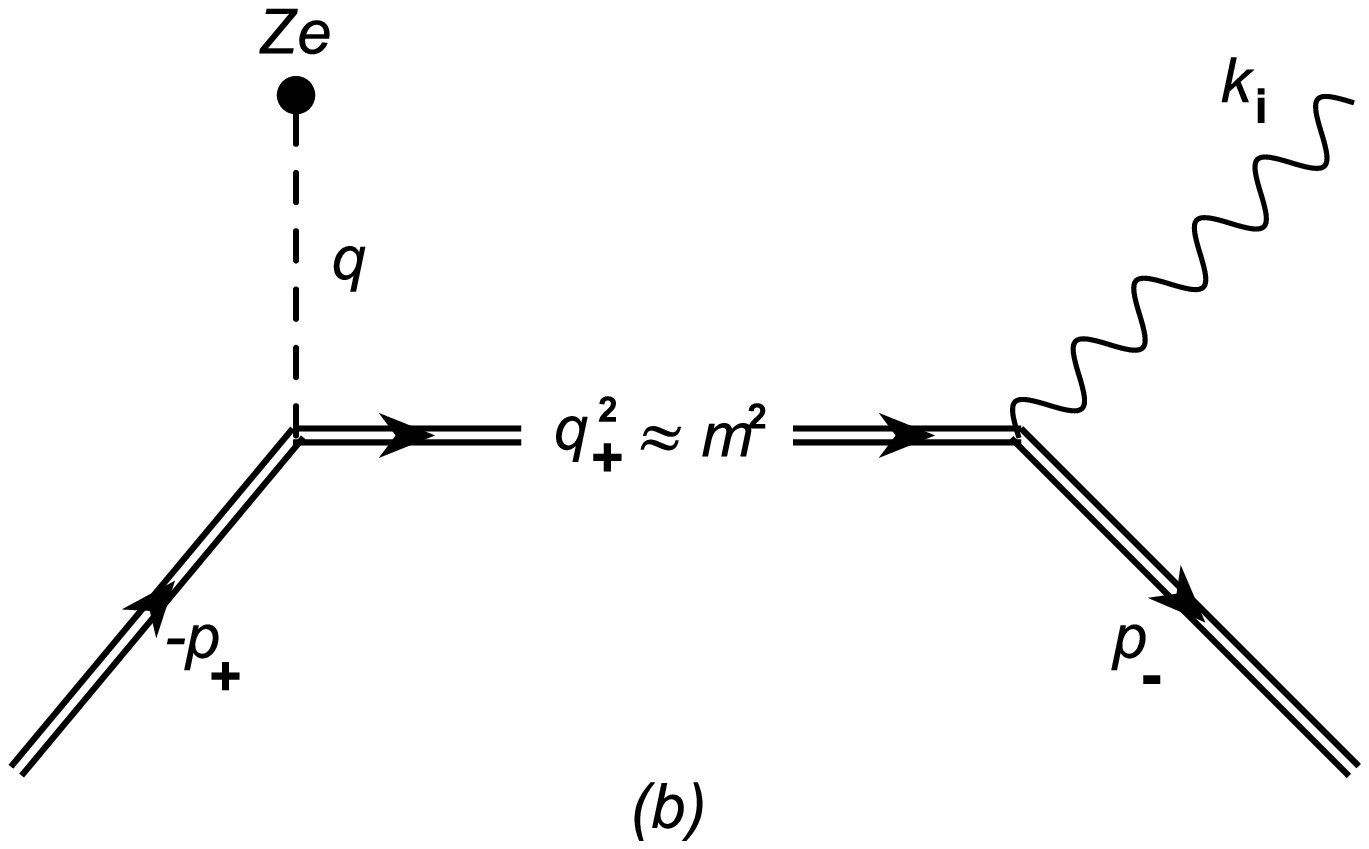}%
\caption{Resonant photoproduction of the electron-positron pair in the field of a nucleus and a pulsed electromagnetic wave.\label{Fig2}}%
\end{figure}

Resonant behavior of the amplitudes (\ref{6})-(\ref{8}), (\ref{23}), (\ref{24}) is explained by quasi-discrete structure: an electron (a positron) $+$ a plane electromagnetic wave. It follows that due to the approximate fulfillment of the energy-momentum conservation law, four-momentum of the intermediate electron (positron) lies near the mass shell.
The following conditions take place for the channels A and B in the resonance \cite{monpul,intech} (see (\ref{23}), (\ref{24}), and also Fig.\ref{Fig2}):
\begin{eqnarray}
\left| {{\beta _ \mp }} \right| \mathbin{\lower.3ex\hbox{$\buildrel<\over
{\smash{\scriptstyle\sim}\vphantom{_x}}$}} 1 \Rightarrow \frac{{\left| {q_ \mp ^2 - {m^2}} \right|}}{{4\left( {k{q_ \mp }} \right)}} \mathbin{\lower.3ex\hbox{$\buildrel<\over
{\smash{\scriptstyle\sim}\vphantom{_x}}$}} \frac{1}{{\omega \tau }} \ll 1.
\label{28}
\end{eqnarray}

The resonant parameters ${\beta _ \mp }$ (\ref{22}) for the case of a weak field (\ref{5}) and the kinematic conditions (\ref{25})-(\ref{27}) have the following form:
\begin{eqnarray}
{\beta _ \mp } = r\frac{{\left[ {4\left( {1 - {x_ \pm }} \right){x_ \pm }{\varepsilon _r} - 4x_ \pm ^2\delta _ \pm ^2 - 1} \right]}}{{16{\varepsilon _r}\left( {1 - {x_ \pm }} \right){x_ \pm }}}\omega \tau,
\label{29}
\end{eqnarray}
\begin{eqnarray}
{\delta _ \pm } = \frac{{{\omega _i}{\theta _{i \pm }}}}{{2m}},\quad{x_ \pm } = \frac{{{E_ \pm }}}{{{\omega _i}}},
\label{30}
\end{eqnarray}
\begin{eqnarray}
{\varepsilon _r} = \frac{{{\omega _i}}}{{{\omega _{thr\left( r \right)}}}},\quad{\omega _{thr\left( r \right)}} = \frac{{{m^2}}}{{r\omega \sin \left( {{{{\theta _i}} \mathord{\left/
 {\vphantom {{{\theta _i}} 2}} \right.
 \kern-\nulldelimiterspace} 2}} \right)}}.
\label{31}
\end{eqnarray}
Herein $r = 1,2,3...$ is a number of a resonance (the number of wave photons, which are absorbed in the laser-stimulated Breit-Wheeler process), ${\omega _{thr\left( r \right)}}$ is the threshold energy for the $r^{th}$-resonance (see (\ref{33})-(\ref{35})). This energy is defined by the rest energy of an electron, the total energy of absorbed wave photons and the angle between the momenta of the initial gamma quantum and a laser wave. Within the optical frequency range, the threshold energy for the first resonance is of the order of ${\omega _{thr\left( 1 \right)}} \sim {10^2}\:{\rm{ GeV}}$. With an increase in the number of a resonance (absorbed photons of a wave), the threshold energy decreases by a factor of $r$ (\ref{31}). In the expression (\ref{31}) we denoted the energy of initial gamma quantum in the units of a threshold energy of $r^{th}$-resonance as ${\varepsilon _r}$. We underline that in the article \cite{Larin} there was studied the first resonance $\left( {r = 1} \right)$ of the process of the ultrarelativistic electron-positron pairs photoproduction in the field of a nucleus and a weak monochromatic laser wave. Wherein, the resonant infinity was eliminated by the Breit-Wigner procedure with the introduction of a radiation width. In the present article, we study the process in the field of a pulsed laser wave for the several first resonances. Herewith, the resonant width, associated with the pulsed nature of a wave, arises as a consequence of the used mathematical apparatus (see (\ref{61})-(\ref{63}), (\ref{87}), (\ref{88})).

\begin{figure}[ht]
\includegraphics[width=1.1\linewidth]{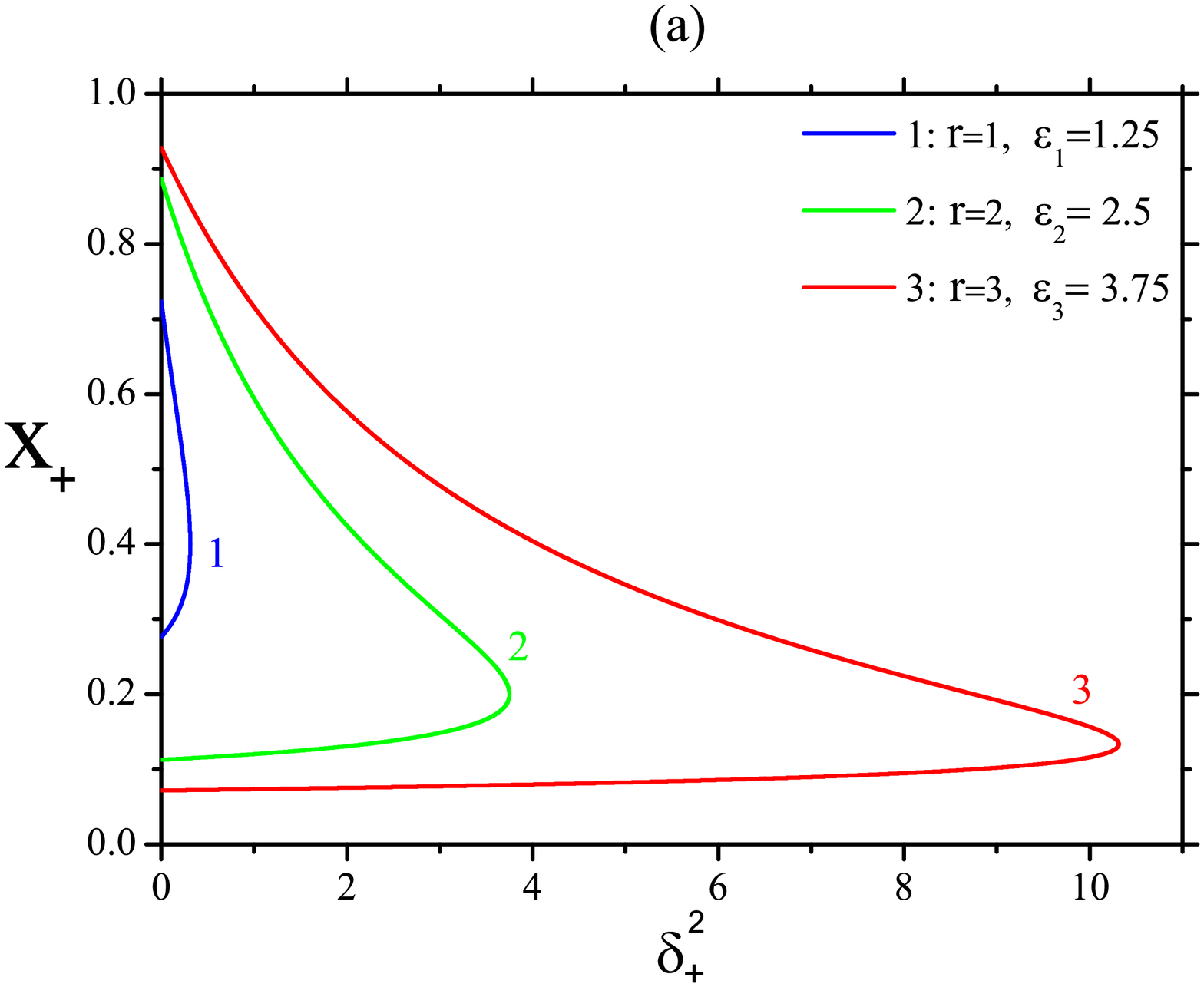}%
\vfill
\includegraphics[width=1.1\linewidth]{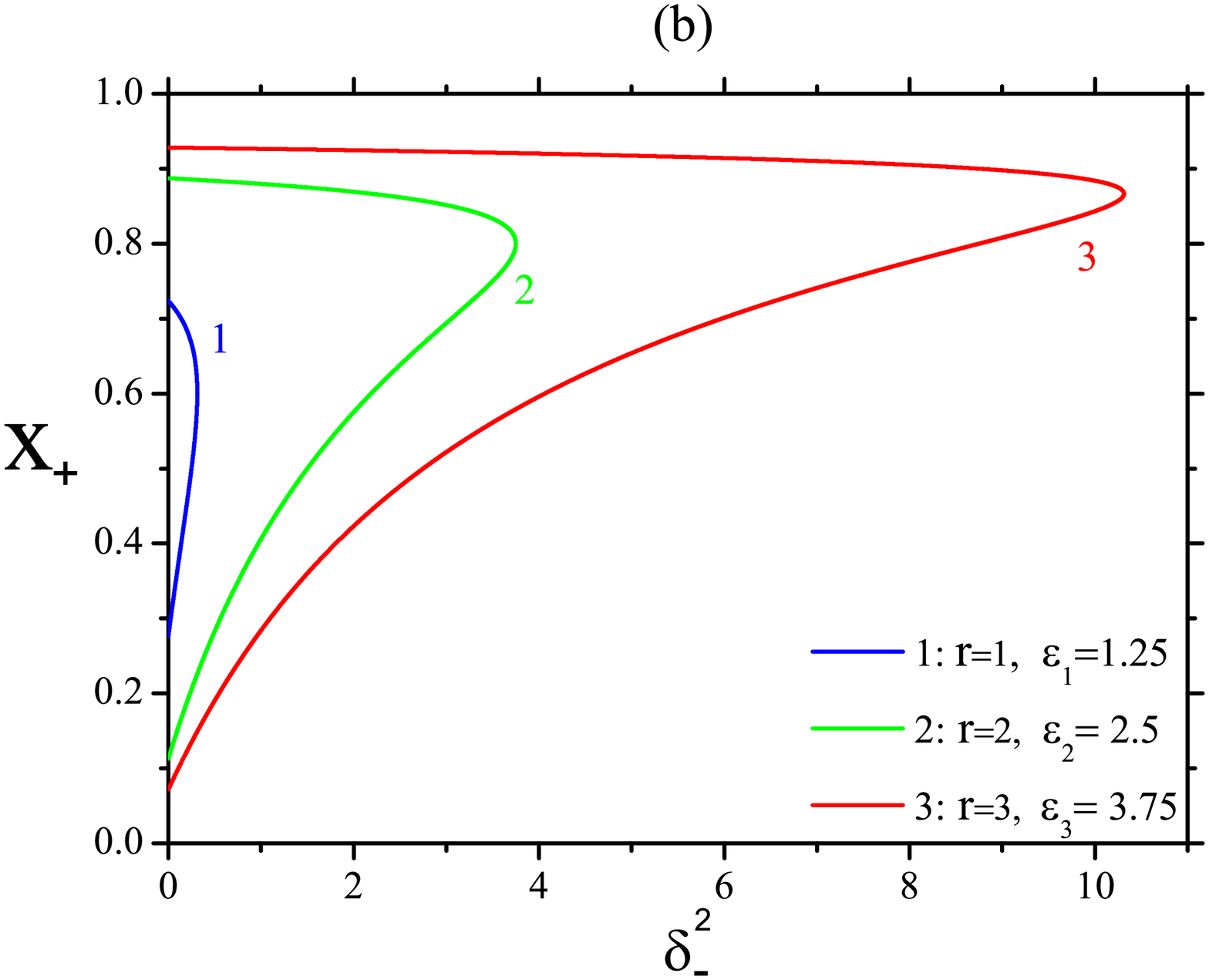}%
\caption{Dependence of the positron resonant energy on the parameter $\delta _ + ^2$ (\ref{33}) (Fig.3a, for the channel A) and on the parameter $\delta _ - ^2$ (\ref{34}) (Fig.3b, for the channel B) for the case ${\omega _i} > {\omega _{thr\left( 1 \right)}}$ and the first three resonances $\left( {{\omega _i} = 125\:{\rm{ GeV }},\:{\omega _{thr\left( 1 \right)}} = 100\:{\rm{ GeV}}} \right)$.\label{Fig3}}%
\end{figure}

\begin{figure}[hb]
\includegraphics[width=1.1\linewidth]{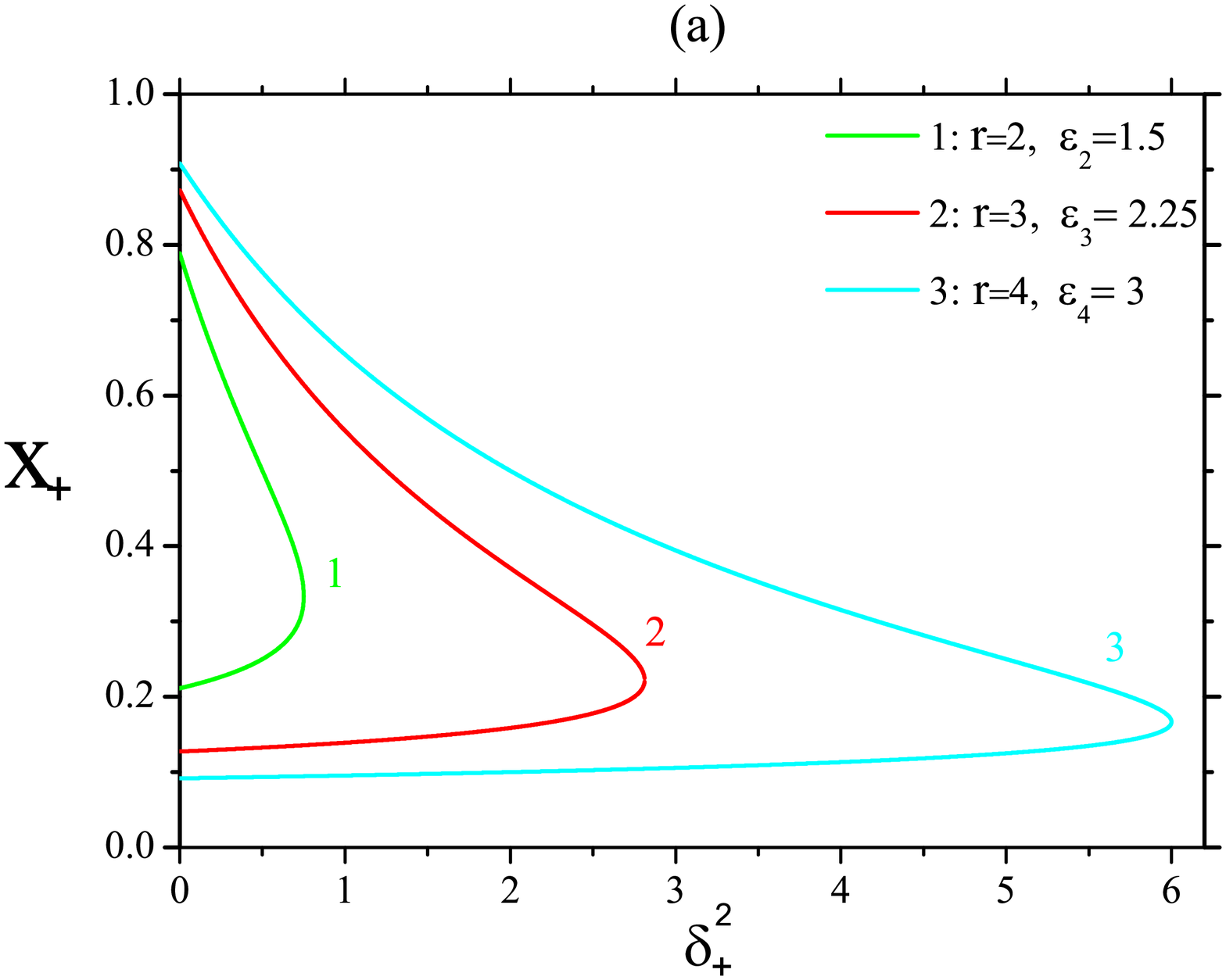}%
\vfill
\includegraphics[width=1.1\linewidth]{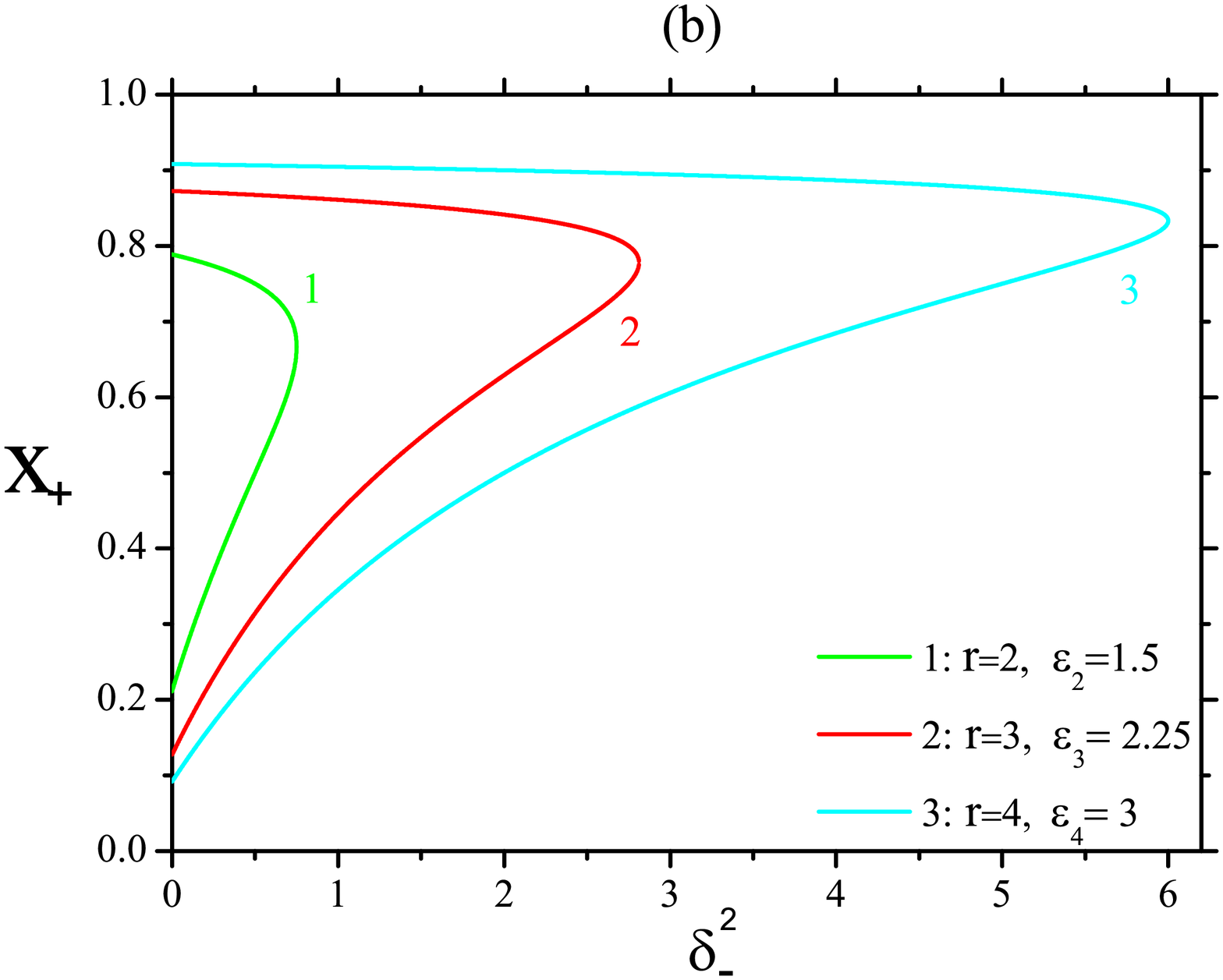}%
\caption{Dependence of the positron resonant energy on the parameter $\delta _ + ^2$ (\ref{33}) (Fig.4a, channel A) and on the parameter $\delta _ - ^2$ (\ref{34}) (Fig.4b, channel B) for the case ${\omega _{thr\left( 1 \right)}} > {\omega _i} > {\omega _{thr\left( 2 \right)}}$ and for the second, third and fourth resonances $\left( {{\omega _i} = 75\:{\rm{ GeV }},\:\omega _{thr(2)}^{} = 50\:{\rm{ GeV}}} \right)$.\label{Fig4}}%
\end{figure}

\begin{figure}[hb]
\includegraphics[width=1.1\linewidth]{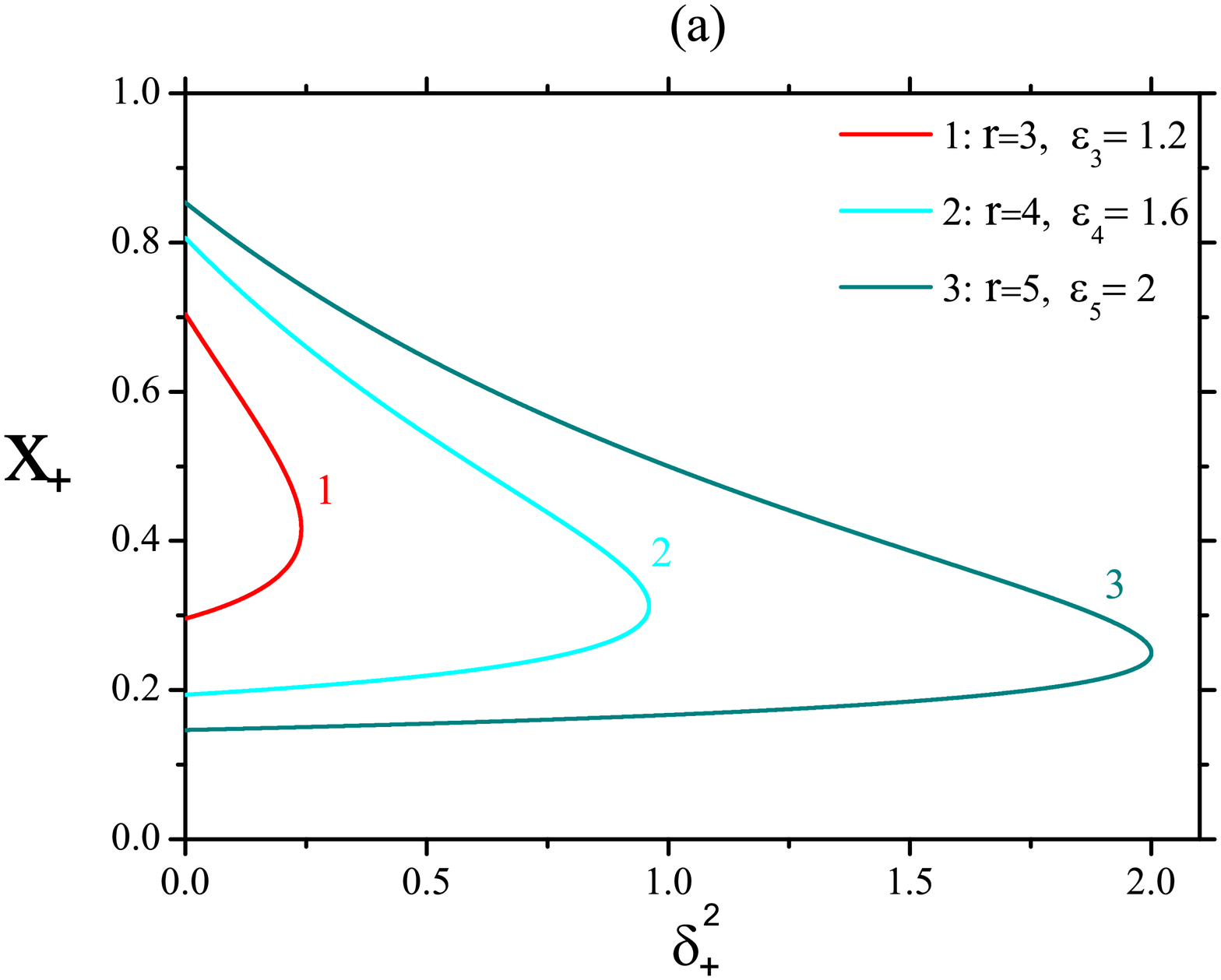}%
\vfill
\includegraphics[width=1.1\linewidth]{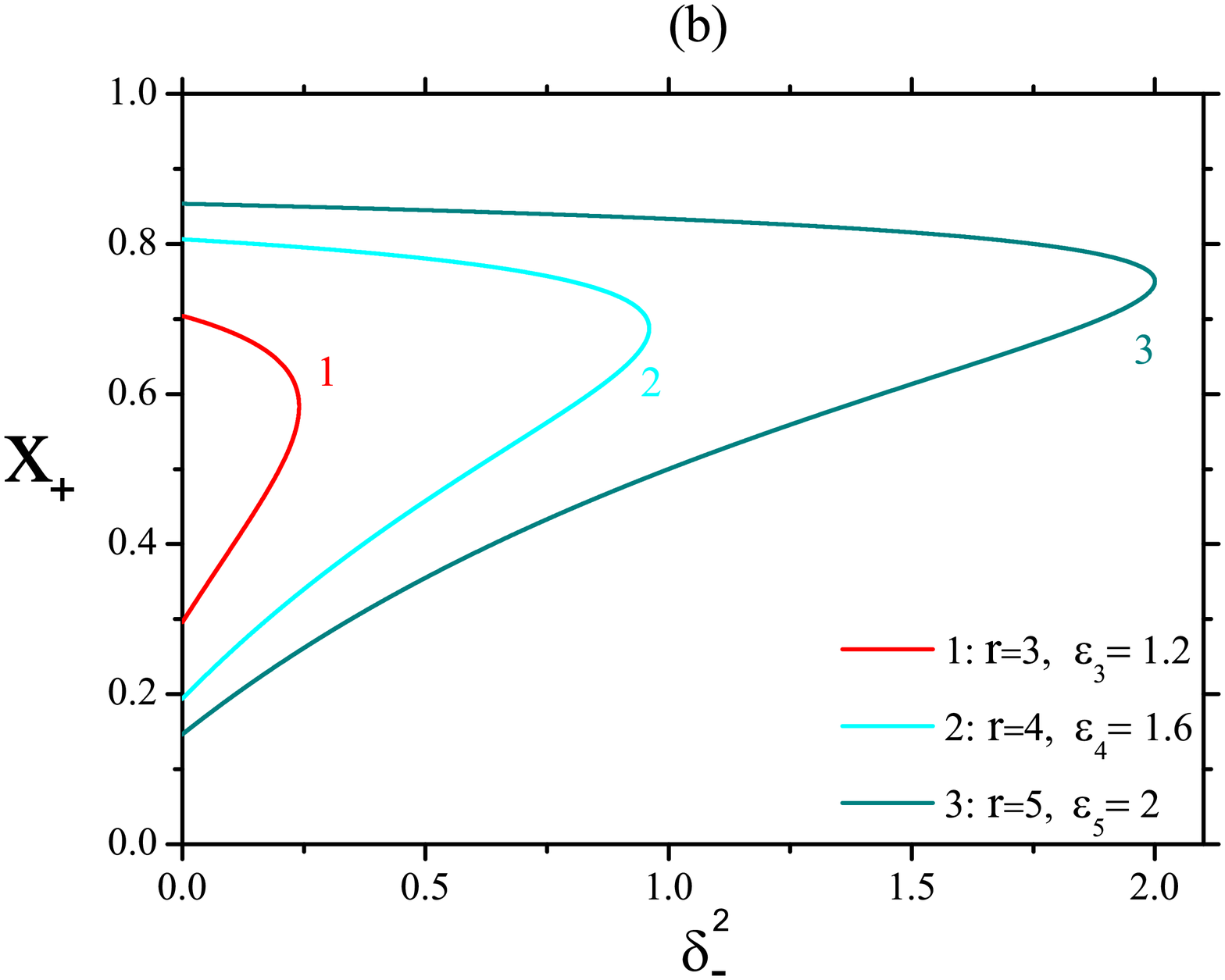}%
\caption{Dependence of the positron resonant energy on the parameter $\delta _ + ^2$ (\ref{33}) (Fig.5a, channel A) and on the parameter $\delta _ - ^2$ (\ref{34}) (Fig.5b, channel B) for the case ${\omega _{thr\left( 2 \right)}} > {\omega _i} > {\omega _{thr\left( 3 \right)}}$ for the third, fourth and fifth resonances $\left( {{\omega _i} = 40\:{\rm{ GeV }},\:\omega _{thr(3)}^{} = 33.3\:{\rm{ GeV}}} \right)$.\label{Fig5}}%
\end{figure}

Using the expressions for the resonant parameters (\ref{29})  we rewrite the resonant conditions (\ref{28}) for the channels A and B:
\begin{eqnarray}
\frac{{r\left| {4\left( {1 - {x_ \pm }} \right){x_ \pm }{\varepsilon _r} - 4x_ \pm ^2\delta _ \pm ^2 - 1} \right|}}{{16{\varepsilon _r}\left( {1 - {x_ \pm }} \right){x_ \pm }}} \mathbin{\lower.3ex\hbox{$\buildrel<\over
{\smash{\scriptstyle\sim}\vphantom{_x}}$}} \frac{1}{{\omega \tau }} \ll 1.
\label{32}
\end{eqnarray}
We can see that the expressions in the modulus have to be close to zero. Hence, we have quadratic equations that define the resonant energies of a positron and an electron. The solutions of these equations for the channels A and B have the following form:
\begin{eqnarray}
{x_{ + (r)}}\left( {\delta _ + ^2} \right) \approx \frac{{{\varepsilon _r} \pm \sqrt {{\varepsilon _r}\left( {{\varepsilon _r} - 1} \right) - \delta _ + ^2} }}{{2\left( {{\varepsilon _r} + \delta _ + ^2} \right)}}{\rm{,  }}\nonumber\\
{x_{ - (r)}}\left( {\delta _ + ^2} \right) \approx 1 - {x_{ + (r)}}\left( {\delta _ + ^2} \right){\rm{,}}
\label{33}
\end{eqnarray}
\begin{eqnarray}
{x_{ - (r)}}\left( {\delta _ - ^2} \right) \approx \frac{{{\varepsilon _r} \pm \sqrt {{\varepsilon _r}\left( {{\varepsilon _r} - 1} \right) - \delta _ - ^2} }}{{2\left( {{\varepsilon _r} + \delta _ - ^2} \right)}}{\rm{, }}\nonumber\\
{x_{ + (r)}}\left( {\delta _ - ^2} \right) \approx 1 - {x_{ - (r)}}\left( {\delta _ - ^2} \right),\quad {\rm{   }}{x_{ \pm (r)}} = \frac{{{E_{ \pm (r)}}}}{{{\omega _i}}}.
\label{34}
\end{eqnarray}
We can conclude that the resonant energies of a positron and an electron for each of the two channels depend on the parameter ${\varepsilon _r}$ and take two different values for a certain positron (the channel A) or electron (the channel B) outgoing angle (see Fig.\ref{Fig3}-Fig.\ref{Fig5}). Herewith, the parameter ${\varepsilon _r}$ has to satisfy the condition:
\begin{eqnarray}
{\varepsilon _r} \ge 1  \to  {\omega _i} \ge {\omega _{thr\left( r \right)}},\quad r = 1,2,3, \ldots\:.
\label{35}
\end{eqnarray}
It follows that value ${\omega _{thr\left( r \right)}}$ (\ref{31}) is the threshold energy for the initial gamma quantum. This threshold energy decreases with an increase in the number of absorbed wave photons. From the expressions (\ref{33}) and (\ref{34}) we can see that the possible outgoing angles depend on the value of the parameter ${\varepsilon _r}$ and are enclosed in the interval:
\begin{eqnarray}
0 \le \delta _ \pm ^2 \le \delta _{\max }^2,\quad \delta _{\max }^2 = {\varepsilon _r}\left( {{\varepsilon _r} - 1} \right),\quad {\varepsilon _r} \ge 1.
\label{36}
\end{eqnarray}

In Fig.\ref{Fig3}a and Fig.\ref{Fig3}b we represent possible values of the positron energy for the channels A and B for the first, second and third resonances in the case of the initial gamma quantum energy is greater than the threshold energy for the first resonance ${\omega _i} > {\omega _{thr\left( 1 \right)}}$. In Fig.\ref{Fig4}a and Fig.\ref{Fig4}b we represent possible values of the positron energy for the channels A and B for the second, third and fourth resonances in the case when the initial gamma quantum energy is greater than the threshold energy for the second resonance ${\omega _{thr\left( 1 \right)}} > {\omega _i} > {\omega _{thr\left( 2 \right)}}$. In Fig.\ref{Fig5}a and Fig.\ref{Fig5}b we represent possible values of the positron energy for the channels A and B for the third, fourth and fifth resonances in the case when the initial gamma quantum energy is greater than the threshold energy for the third resonance ${\omega _{thr\left( 2 \right)}} > {\omega _i} > {\omega _{thr\left( 3 \right)}}$. These figures show that the positron energy for every outgoing angle ($\delta _ + ^2$ for channel A and $\delta _ - ^2$  for channel B) can take two values. Moreover, the most difference in the maximum and minimum values of the energy takes place at zero outgoing angle of a positron (electron) relative to the initial gamma quantum momentum $\left( {\delta _ \pm ^2 = 0} \right)$. With an increase in the outgoing angle of a positron (electron) the difference between the maximum and minimum values of the positron energy decreases and for the supreme outgoing angle ($\delta _ \pm ^2 = \delta _{\max }^2$ (\ref{36})) the positron energy takes a single value (\ref{33}), (\ref{34}). Note that similar reasoning holds true for the electron.

It is very important to emphasize that the electron-positron pair energies for the channel A depend only on the outgoing angle of a positron (parameter $\delta _ + ^2$) and for the channel B depend only on the outgoing angle of an electron (parameter $\delta _ - ^2$). Accordingly, the channels A and B do not interfere with each other. Besides, in the frame of one of the channels (A or B) different resonances also does not interfere (see Fig.\ref{Fig3}-Fig.\ref{Fig5}, there are no intersections of the curves with different $r$). Due to that fact we can ignore the interference of the channels A and B and also interference between different resonances in the frame of one channel, when calculating the resonant differential cross section. Herewith, we will exclude the case of the electron (positron) scattering at a zero angle $\left( {\delta _ - ^2 = \delta _ + ^2 = 0} \right)$.

\section{RESONANT CROSS SECTION OF THE PPP}
We simplify amplitudes of the process (\ref{6})-(\ref{8}), (\ref{23}), (\ref{24}) for the case of the resonant kinematics (\ref{25})-(\ref{27}), (\ref{33}), (\ref{34}). Consequently, the arguments of the Bessel function (\ref{16}) in the amplitudes $M_{(l + r)}^0$ (\ref{13}) and $F_{( - r)}^\mu $ (\ref{14}) are small $\left( {{\gamma _{p'p}} \mathbin{\lower.3ex\hbox{$\buildrel<\over
{\smash{\scriptstyle\sim}\vphantom{_x}}$}} {\eta _0} \ll 1} \right)$. So we can put:
\begin{eqnarray}
M_{(l + r)}^0\left( {p',p,{\phi _2}} \right) \approx {\tilde \gamma ^0}{L_0}\left( {p',p,{\phi _2}} \right) \approx {\tilde \gamma ^0}{\rm{  }}\left( {l =  - r} \right),
\label{37}
\end{eqnarray}
\begin{eqnarray}
F_{( - r)}^\mu \left( {p',p,{\phi _1}} \right) = {\tilde \gamma ^\mu }{L_{- r}}\left( {p',p,{\phi _1}} \right) +{L_{- r+1}}\left( {p',p,{\phi _n}} \right)\eta \left( {{\phi _1}} \right)\times\nonumber\\
\times\left[ {\frac{m}{{4\left( {kp'} \right)}}{{\hat e}_ + }\hat k{{\tilde \gamma }^\mu } - \frac{m}{{4\left( {kp} \right)}}{{\tilde \gamma }^\mu }\hat k{{\hat e}_ + }} \right].\quad\quad
\label{38}
\end{eqnarray}
We will carry out further analysis for a certain type of the envelope function. Let us choose the envelope function in a Gaussian-like form:
\begin{eqnarray}
g\left( {{\phi _1}} \right) = \exp \left\{ { - {{\left( {2{\phi _1}} \right)}^2}} \right\}.
\label{39}
\end{eqnarray}
We apply series expansion for the Bessel function with respect to the small parameter in the amplitudes (\ref{38}) and also integrate with respect to ${\phi _1}$. Finally, the amplitude of the PPP for the resonance with number $r$ takes the following form:
\begin{eqnarray}
{S_{( - r)}} = \frac{{{\pi ^2}{\tau ^2}\omega Z{e^3}}}{{\sqrt {2{\omega _i}{E_ - }{E_ + }} }}\frac{{\left( {{{\bar u}_{{p_ - }}}{B_{(r)}}{v_{ - {p_ + }}}} \right)}}{{\left[ {{{\bf{q}}^2} + {q_0}\left( {{q_0} - 2{q_z}} \right)} \right]}},
\label{40}
\end{eqnarray}
where
\begin{eqnarray}
{B_{(r)}} = {B_{ + \left( r \right)}} + {B_{ - \left( r \right)}}.
\label{41}
\end{eqnarray}
Here ${B_{ + \left( r \right)}}$ and ${B_{ - \left( r \right)}}$ are the resonant amplitudes for the channels A and B
\begin{eqnarray}
{B_{ + (r)}} = {U_{(r)}}\left( {{q_ - }} \right)\left[ {{{\tilde \gamma }^0}\left( {{{\hat q}_ - } + m} \right)\left( {{\varepsilon _\mu }z_{ + (r)}^\mu } \right)} \right],
\label{42}
\end{eqnarray}
\begin{eqnarray}
{B_{ - (r)}} = {U_{(r)}}\left( {{q_ + }} \right)\left[ {{{\tilde \gamma }^0}\left( {{{\hat q}_ + } + m} \right)\left( {{\varepsilon _\nu }z_{ - (r)}^\nu } \right)} \right],
\label{43}
\end{eqnarray}
\begin{eqnarray}
{U_{(r)}}\left( {{q_ \pm }} \right) = \frac{{\eta _0^r}}{{\sqrt r \left( {k{q_ \pm }} \right)}}\exp \left( { - \frac{{\beta _ \pm ^2}}{{4r}}} \right){V_r}\left( {{q_0},{\beta _ \pm }} \right),
\label{44}
\end{eqnarray}
\begin{eqnarray}
{V_r}\left( {{q_0},{\beta _ \pm }} \right) = \int\limits_{ - \infty }^\infty  {d{\phi _2}} \exp \left[ {i\left( {{q_0}\tau  + 2{\beta _ \pm }} \right){\phi _2}} \right]\times\nonumber\\
\times erf\left( {2\sqrt r \phi _2^{} + \frac{{i{\beta _ \pm }}}{{2\sqrt r }}} \right).
\label{45}
\end{eqnarray}
We point out that there is no summation over $r$ in the amplitudes (\ref{42}) and (\ref{43}) due to the absence of the interference of resonances with different $r$. The expressions $z_{ \pm (r)}^\mu $ have the following form:
\begin{eqnarray}
z_{ + (r)}^\mu  = {\tilde \gamma ^\mu }c_ + ^r + c_ + ^{r - 1}{\eta _0}\left[ {\frac{m}{{4\left( {k{q_ - }} \right)}}{{\hat e}_ + }\hat k{{\tilde \gamma }^\mu } - \frac{m}{{4\left( {k{p_ + }} \right)}}{{\tilde \gamma }^\mu }\hat k{{\hat e}_ + }} \right],\quad\quad
\label{46}
\end{eqnarray}
\begin{eqnarray}
z_{ - (r)}^\nu  = {\tilde \gamma ^\mu }c_ - ^r + c_ - ^{r - 1}{\eta _0}\left[ {\frac{m}{{4\left( {k{q_ + }} \right)}}{{\hat e}_ + }\hat k{{\tilde \gamma }^\mu } - \frac{m}{{4\left( {k{p_ - }} \right)}}{{\tilde \gamma }^\mu }\hat k{{\hat e}_ + }} \right].\quad\quad
\label{47}
\end{eqnarray}
Herein we introduced the following notation:
\begin{eqnarray}
c_ \pm ^r = \frac{{{{\left( { - 1} \right)}^r}\gamma _ \pm ^r}}{{r!}}\exp \left( {ir{\chi _{{q_ \mp }{p_ \pm }}}} \right),
\label{48}
\end{eqnarray}
\begin{eqnarray}
{\gamma _ \pm } = r\sqrt {\frac{{{u_ \pm }}}{{{u_r}}}\left( {1 - \frac{{{u_ \pm }}}{{{u_r}}}} \right)}.
\label{49}
\end{eqnarray}
In the expression (\ref{49}) the relativistically invariant parameters for the channels A and B are defined in the following way \cite{Ritus}:
\begin{eqnarray}
{u_ \pm } = \frac{{{{\left( {k{k_i}} \right)}^2}}}{{4\left( {k{p_ \pm }} \right)\left( {k{q_ \mp }} \right)}},\quad{\rm{   }}{u_r} = r\frac{{\left( {k{k_i}} \right)}}{{2{m^2}}}.
\label{50}
\end{eqnarray}
We want to emphasize that the applicability of the obtained expansions for the Bessel functions (\ref{16}) is valid when the following inequality holds for the number of resonance and wave intensity:
\begin{eqnarray}
\frac{r}{{\sqrt {r + 1} }} \ll \eta _0^{ - 1}.
\label{51}
\end{eqnarray}

We obtained the probability of the resonant process for the entire observation time in the interference absence of the channels A and B (first and second terms in (\ref{41})):
\begin{eqnarray}
dw_{ \pm (r)}^{} = \frac{{{\pi ^4}{\omega ^2}{\tau ^4}{Z^2}{e^6}}}{{2{\omega _i}{E_ + }{E_ - }}}\frac{{{{\left| {{{\bar u}_{{p_ - }}}{B_{ \pm (r)}}{v_{ - {p_ + }}}} \right|}^2}}}{{{{\left[ {{{\bf{q}}^2} + {q_0}\left( {{q_0} - 2{q_z}} \right)} \right]}^2}}}\frac{{{d^3}{p_ + }{d^3}{p_ - }}}{{T{{\left( {2\pi } \right)}^6}}}.\quad
\label{52}
\end{eqnarray}
Herein $T$ is a some relatively large observation time $\left( {T \mathbin{\lower.3ex\hbox{$\buildrel>\over
{\smash{\scriptstyle\sim}\vphantom{_x}}$}} \tau } \right)$. The differential cross section of the PPP process on a nucleus in the field of a pulsed light wave is obtained from the probability per unit of time by dividing it by density of the incident particles flux. The flux is equal to unity for the case of incident gamma quanta. Then, averaging over the initial gamma quanta polarizations and summation over polarizations of the final electrons and positrons is carried out in the standard way \cite{LL}. Thus, after all the calculations, we derived the resonant differential cross section of the PPP (for example for the channel A) in the following form: 
\begin{eqnarray}
d\sigma _{ + (r)}^{} = \frac{{\pi \tau {{\left( {\omega \tau } \right)}^2}}}{{4{{\left( {2\pi } \right)}^3}}}\left( {\frac{\tau }{T}} \right)\frac{{{m^2}\left| {{{\bf{q}}_ - }} \right|}}{{r{{\left( {k{q_ - }} \right)}^2}}}\exp \left( { - \frac{{\beta _ - ^2}}{{2r}}} \right)d{W_{ + (r)}}\left( {{q_ - },{p_ + }}\right)\times\nonumber\\
\times\frac{{2{Z^2}r_e^2{m^2}\left[ {{m^2} + p_ - ^0q_ - ^0 + {{\bf{p}}_ - }{{\bf{q}}_ - }} \right]}}{{{E_ - }\left| {{{\bf{q}}_ - }} \right|{{\left[ {{{\bf{q}}^2} + {q_0}\left( {{q_0} - 2{q_z}} \right)} \right]}^2}}}{\left| {{V_r}\left( {{q_0},{\beta _ - }} \right)} \right|^2}{d^3}{p_ - },\quad\quad
\label{53}
\end{eqnarray}
where $d{W_{ + (r)}}\left( {{q_ - },{p_ + }} \right)$
is a probability per unit of time of the electron-positron pair photoproduction with the momenta ${q_ - }$ and ${p_ + }$ , correspondingly, due to absorption of $r$-photons of a wave  (the laser-stimulated Breit-Wheeler process) \cite{Ritus}:
\begin{eqnarray}
d{W_{ + (r)}}\left( {{q_ - },{p_ + }} \right) = \frac{\alpha }{{{\omega _i}{E_ + }}}\eta _0^{2r}{\left( {\frac{{{r^r}}}{{r!}}} \right)^2}{\left[ {\frac{{{u_ + }}}{{{u_r}}}\left( {1 - \frac{{{u_ + }}}{{{u_r}}}} \right)} \right]^{r - 1}}\times\nonumber\\
\times\left[ {2{u_ + } - 1 + 2\frac{{{u_ + }}}{{{u_r}}}\left( {1 - \frac{{{u_ + }}}{{{u_r}}}} \right)} \right]{d^3}{p_ + }.\quad\quad
\label{54}
\end{eqnarray}
Here $\alpha $ is the fine structure conastant and the relativistically invariant parameters ${u_ + },\;{u_r}$ are defined by the expressions (\ref{50}). We integrate the expression (\ref{53}) with respect to the final electron energy ${d^3}{p_ - } = \left| {{{\bf{p}}_ - }} \right|{E_ - }d{E_ - }d{\Omega _ - }$.  In order to deduce the corresponding differential cross section for the channel B, it is necessary to integrate with respect to the final positron energy ${d^3}{p_ + } = \left| {{{\bf{p}}_ + }} \right|{E_ + }d{E_ + }d{\Omega _ + }$. Herewith, we make the change of the integration variable:
\begin{eqnarray}
\zeta  = \frac{{{q_0}}}{\omega } = \frac{1}{\omega }\left( {{E_ + } + {E_ - } - {\omega _i} - r\omega } \right),\quad d{E_ - } = \omega d\zeta.
\label{55}
\end{eqnarray}
It is easy to see that the main contribution to the integral (\ref{53}) is made by values from the interval
\begin{eqnarray}
\left| \zeta  \right| \mathbin{\lower.3ex\hbox{$\buildrel<\over
{\smash{\scriptstyle\sim}\vphantom{_x}}$}} \frac{1}{{\omega \tau }} \ll 1.
\label{56}
\end{eqnarray}
Therefore, in the numerator of the expression (\ref{53}) we may put $\zeta  = 0$ everywhere but the exponent in ${V_r}\left( {\omega \zeta ,{\beta _ - }} \right)$ (\ref{45}). In addition, since the transferred momentum, which makes the main contribution to the resonant cross section, has the order of magnitude ${{\bf{q}}^2} \sim {m^4}/\omega _i^2 \mathbin{\lower.3ex\hbox{$\buildrel<\over
{\smash{\scriptstyle\sim}\vphantom{_x}}$}} {m^2}\omega /{\omega _i}$ \cite{Larin} and the correction to its value is much less $\left| {{q_z}} \right|\omega \zeta  \sim {m^2}\omega \zeta /{\omega _i} \mathbin{\lower.3ex\hbox{$\buildrel<\over
{\smash{\scriptstyle\sim}\vphantom{_x}}$}} {m^2}\omega /\left( {{\omega _i}\omega \tau } \right) \ll {m^2}\omega /{\omega _i}$, then in the denominator (\ref{53}) we can neglect corrections to the square of the transferred momentum ${{\bf{q}}^2}$. Given this, it is easy to integrate the expression (\ref{53}) with respect to $\zeta $. The resonant differential cross sections of the PPP for the channels A and B take the form:
\begin{eqnarray}
d{\sigma _{ + (r)}} = d{\sigma _{(0)}}\left( {{p_ - },{q_ - }} \right)\Psi _{(r)}^{res}\left( {{q_ - }} \right)d{W_{ + (r)}}\left( {{q_ - },{p_ + }} \right),
\label{57}
\end{eqnarray}
\begin{eqnarray}
d{\sigma _{ - (r)}} = d{\sigma _{(0)}}\left( {{p_ + },{q_ + }} \right)\Psi _{(r)}^{res}\left( {{q_ + }} \right)d{W_{ - (r)}}\left( {{q_ + },{p_ - }} \right).
\label{58}
\end{eqnarray}
Here $d{W_{ + (r)}}\left( {{q_ - },{p_ + }} \right)$ is defined by the expression (\ref{54}), $d{W_{ - (r)}}\left( {{q_ + },{p_ - }} \right)$ is obtained from (\ref{54}) by corresponding replacement of the four-momenta, $d{\sigma _{(0)}}\left( {{p_ - },{q_ - }} \right)$ and $d{\sigma _{(0)}}\left( {{p_ + },{q_ + }} \right)$ are the differential cross sections of the intermediate electron (positron) scattering on a nucleus in the absence of radiation (absorption) of photons \cite{Fedorov}: 
\begin{eqnarray}
d{\sigma ^{\left( 0 \right)}}\left( {{p_ - },{q_ - }} \right) = 2{Z^2}r_e^2\frac{{\left| {{{\bf{p}}_ - }} \right|}}{{\left| {{{\bf{q}}_ - }} \right|}}\frac{{{m^2}\left[ {{m^2} + p_ - ^0q_ - ^0 + {{\bf{p}}_ - }{{\bf{q}}_ - }} \right]}}{{{{\bf{q}}^4}}}d{\Omega _ - },\nonumber\\
{\rm{  }}{\bf{q}} = {{\bf{p}}_ - } - {{\bf{q}}_ - }.\quad\quad
\label{59}
\end{eqnarray}
Herein $d{\sigma _{(0)}}\left( {{p_ + },{q_ + }} \right)$ can be obtained from the expression (\ref{59}) by replacement of four-momenta: ${q_ - } \to {q_ + },{\rm{ }}{p_ - } \to {p_ + }$. The resonant functions $\Psi _{(r)}^{res}\left( {{q_ \pm }} \right)$  have the following forms:
\begin{eqnarray}
\Psi _{res}^{\left( r \right)}\left( {{q_ \pm }} \right) = \frac{{{m^2}{{\left( {\omega \tau } \right)}^2}\left| {{{\bf{q}}_ \pm }} \right|}}{{16{\pi ^2}r{{\left( {k{q_ \pm }} \right)}^2}}}P_{(r)}^{res}\left( {{\beta _ \pm }} \right),
\label{60}
\end{eqnarray}
where we introduced a function of resonant profile  $P_{(r)}^{res}\left( {{\beta _ \pm }} \right)$ which depends on the resonant parameter ${\beta _ \pm }$ and has the following expression:
\begin{eqnarray}
P_{(r)}^{res}\left( {{\beta _ \pm }} \right) = \exp \left( { - \frac{{\beta _ \pm ^2}}{{2r}}} \right)\frac{1}{{2\rho }}\int\limits_{ - \rho }^\rho  {d{\phi _2}} {\left| {erf\left( {2\sqrt r \phi _2^{} + \frac{{i{\beta _ \pm }}}{{2\sqrt r }}} \right)} \right|^2}, \nonumber\\
\rho  = {T \mathord{\left/
 {\vphantom {T \tau }} \right.
 \kern-\nulldelimiterspace} \tau }.\quad\quad\quad
\label{61}
\end{eqnarray}
\begin{figure}[ht]
\includegraphics[width=1.1\linewidth]{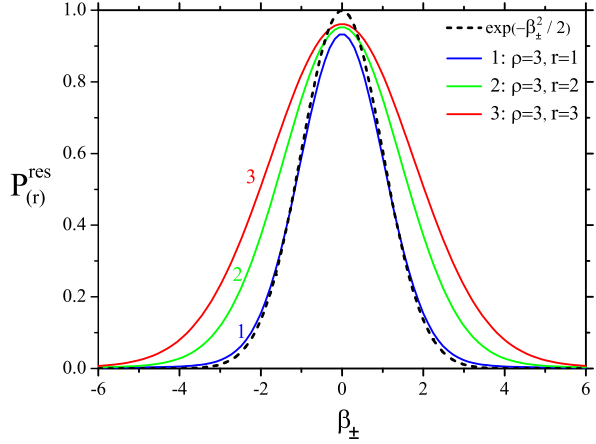}%
\caption{The resonant profile function dependence $P_{(r)}^{res}$ (\ref{61}) on the resonant parameter ${\beta _ \pm }$ for the first three resonances.\label{Fig6}}%
\end{figure}

We are interested in the resonant profile function (\ref{61}) when resonant parameter $\left| {{\beta _ \pm }} \right| \ll 1$. In such case the function of resonant profile $P_{(r)}^{res}\left( {{\beta _j}} \right)$ transforms into the Breit-Wigner formula \cite{monpul,intech}:
\begin{eqnarray}
P_{(r)}^{res}\left( {\left| {{\beta _ \pm }} \right| \ll 1} \right) \approx \frac{{{a_{(r)}}\Lambda _{ \pm \left( r \right)}^2}}{{\left[ {{{\left( {q_ \pm ^2 - m_{}^2} \right)}^2} + \Lambda _{ \pm \left( r \right)}^2} \right]}} \approx P_{\max (r)}^{res} = {a_{(r)}}.\quad
\label{62}
\end{eqnarray}
Herein $\Lambda _{ \pm \left( r \right)}^{}$ is a width of the corresponding resonance, which is determined by the duration of the laser pulse:
\begin{eqnarray}
{\Lambda _{ \pm \left( r \right)}} = \frac{{4r{c_{(r)}}\left( {k{q_ \pm }} \right)}}{{\left( {\omega \tau } \right)}},\quad {c_{(r)}} = \sqrt {\frac{{2a_{(r)}^{}}}{{r\left( {{a_{(r)}} - 2r{b_{(r)}}} \right)}}} ,{\rm{ }}
\label{63}
\end{eqnarray}
\begin{eqnarray}
{a_{(r)}} = \frac{1}{\rho }\int\limits_0^\rho  {er{f^2}\left( {2\sqrt r \phi } \right)d\phi},
\label{64}
\end{eqnarray}
\begin{eqnarray}
\begin{array}{l}
{b_{(r)}} = \frac{1}{{\rho \sqrt {\pi r} }}\left[ {\frac{1}{{2\sqrt {2\pi } r}}erf\left( {2\sqrt {2r} \rho } \right) + } \right.\\
\left. { + 2\int\limits_0^\rho  {\phi  erf\left( {2\sqrt r \phi } \right)\exp \left( { - 4r\phi _{}^2} \right)d\phi } } \right].
\end{array}
\label{65}
\end{eqnarray}
For the first five resonances we have: ${a_{(r)}} \sim 1,{\rm{ }}{c_{(r)}} \sim 1,{\rm{ }}{b_{(r)}} \sim {10^{ - 2}}$. From the relation (\ref{62}) it follows that the maximum of the function is determined by the parameter ${a_{(r)}}$ (\ref{64}), which depends on the number of resonance and the ratio between the observation time and the duration of a laser pulse (the parameter $\rho $). In Fig.\ref{Fig6} we can see that resonances take place if the resonant parameters $\left| {{\beta _ \pm }} \right| \mathbin{\lower.3ex\hbox{$\buildrel<\over
{\smash{\scriptstyle\sim}\vphantom{_x}}$}} 1$. When $\left| {{\beta _ \pm }} \right| \gg 1$ the resonant profile function exponentially decreases \cite{monpul,intech}.

\section{ANALYSIS OF THE RESONANT DIFFERENTIAL CROSS SECTION OF THE PPP BY A HIGH-ENERGY GAMMA QUANTUM}
We rewrite the resonant differential cross section of the PPP (\ref{57})-(\ref{61}) accordingly to the resonant kinematic (\ref{25})-(\ref{27}):
\begin{eqnarray}
d{\sigma _{ + (r)}} = \alpha {Z^2}r_e^2\eta _0^{2r}\frac{{r{{\left( {\omega \tau } \right)}^2}\left( {1 - {x_{ + (r)}}} \right){x_{ + (r)}}}}{{8\pi d_{ + (r)}^2}}\frac{{{K_{ + (r)}}}}{{\varepsilon _r^2}}P_{(r)}^{res}\left( {{\beta _ - }}\right)\times\nonumber\\
\times d\delta _ + ^2d\delta _ - ^2d{x_{ + (r)}}d\varphi, \quad\quad\quad
\label{66}
\end{eqnarray}
\begin{eqnarray}
d{\sigma _{ - (r)}} = \alpha {Z^2}r_e^2\eta _0^{2r}\frac{{r{{\left( {\omega \tau } \right)}^2}\left( {1 - {x_{ - (r)}}} \right){x_{ - (r)}}}}{{8\pi d_{ - (r)}^2}}\frac{{{K_{ - (r)}}}}{{\varepsilon _r^2}}P_{(r)}^{res}\left( {{\beta _ + }} \right)\times\nonumber\\
\times d\delta _ + ^2d\delta _ - ^2d{x_{ - (r)}}d\varphi. \quad\quad\quad
\label{67}
\end{eqnarray}
Herein $\varphi $ is an angle between the planes $\left( {{{\bf{k}}_i}{{\bf{p}}_ + }} \right){\rm{ and }}\left( {{{\bf{k}}_i}{{\bf{p}}_ - }} \right)$. The resonant profile functions $P_{(r)}^{res}\left( {{\beta _ \mp }} \right)$  are determined by the expression (\ref{61}), where the resonant parameters are given by the relation (\ref{29}). Functions ${K_{ \pm (r)}}$  are defined by probability of the laser-stimulated Breit-Wheeler process:
\begin{eqnarray}
{K_{ \pm (r)}} = {\left( {\frac{{{r^r}}}{{r!}}} \right)^2}{\left[ {\frac{{4x_{ \pm (r)}^2\delta _ \pm ^2}}{{{{\left( {1 + 4x_{ \pm (r)}^2\delta _ \pm ^2} \right)}^2}}}} \right]^{r - 1}}\times\nonumber\\
\times \left[ {\frac{{16x_{ \pm (r)}^2\delta _ \pm ^2}}{{{{\left( {1 + 4x_{ \pm (r)}^2\delta _ \pm ^2} \right)}^2}}} + \left( {\frac{{x_{ \pm (r)}^{}}}{{1 - x_{ \pm (r)}^{}}} + \frac{{1 - x_{ \pm (r)}^{}}}{{x_{ \pm (r)}^{}}}} \right)} \right].
\label{68}
\end{eqnarray}
The values ${d_{ \pm (r)}}$  are defined by square of the transferred momentum:
\begin{eqnarray}
{d_{ \pm (r)}} = {d_0}\left( {{x_{ \pm (r)}}} \right) +{\left( {\frac{m}{{2{\omega _i}}}} \right)^2}\times\nonumber\\
\times \left( {g_0^2\left( {{x_{ \pm (r)}}} \right) + \frac{{4{\varepsilon _r}}}{{\sin \left( {{\theta _i}/2} \right)}}\left[ {4{\varepsilon _r} - g_0^2\left( {{x_{ \pm (r)}}} \right)} \right]} \right),
\label{69}
\end{eqnarray}
where
\begin{eqnarray}
{d_0}\left( {{x_{ \pm (r)}}} \right) = \tilde \delta _ + ^2 + \tilde \delta _ - ^2 + 2\tilde \delta _ + ^{}\tilde \delta _ - ^{}\cos \varphi ,{\rm{  }}
\label{70}
\end{eqnarray}
\begin{eqnarray}
{g_0}\left( {{x_{ + (r)}}} \right) = \frac{{1 + \tilde \delta _ + ^2}}{{{x_{ + (r)}}}} + \frac{{1 + \tilde \delta _ - ^2}}{{1 - {x_{ + (r)}}}},\nonumber\\
{\rm{  }}{g_0}\left( {{x_{ - (r)}}} \right) = \frac{{1 + \tilde \delta _ - ^2}}{{{x_{ - (r)}}}} + \frac{{1 + \tilde \delta _ + ^2}}{{1 - {x_{ - (r)}}}},{\rm{  }}
\label{71}
\end{eqnarray}
\begin{eqnarray}
{\tilde \delta _ \pm } = 2{x_{ \pm (r)}}{\delta _ \pm }.
\label{72}
\end{eqnarray}

In the same kinematic region, the differential cross section of the PPP in the absence of an external laser field $d{\sigma _*}$  has the following expression \cite{Larin,LL}:
\begin{eqnarray}
d\sigma _{ * \left(  \pm  \right)}^{} = \frac{{128}}{\pi }{Z^2}r_e^2\alpha x_ \pm ^3{\left( {1 - {x_ \pm }} \right)^3}\times\nonumber\\
\times \frac{{{D_0}\left( {{x_ \pm }} \right) + {{\left( {m/{\omega _i}} \right)}^2}{D_1}\left( {{x_ \pm }} \right)}}{{{{\left[ {{d_0}\left( {{x_ \pm }} \right) + {{\left( {m/2{\omega _i}} \right)}^2}g_0^2\left( {{x_ \pm }} \right)} \right]}^2}}}d\delta _ + ^2d\delta _ - ^2d{x_ \pm }d\varphi. \label{73}
\end{eqnarray}
Here the top sign "+" means that the cross section without a field was integrated with respect to the electron final energy and the only dependence on the positron energy is remained. The bottom sign "-" corresponds to the integration of the cross section with respect to the positron final energy and the dependence on the electron energy is remained.  In the expression  (\ref{73}) we denoted:
\begin{eqnarray}
{D_0}\left( {{x_ \pm }} \right) =  - \frac{{\tilde \delta _ + ^2}}{{{{\left( {1 + \tilde \delta _ + ^2} \right)}^2}}} - \frac{{\tilde \delta _ - ^2}}{{{{\left( {1 + \tilde \delta _ - ^2} \right)}^2}}} +\nonumber\\
+\frac{1}{{2{x_ \pm }(1 - {x_ \pm })}}\frac{{\tilde \delta _ + ^2 + \tilde \delta _ - ^2}}{{\left( {1 + \tilde \delta _ + ^2} \right)\left( {1 + \tilde \delta _ - ^2} \right)}} +\nonumber\\
+\left( {\frac{{{x_ \pm }}}{{1 - {x_ \pm }}} + \frac{{1 - {x_ \pm }}}{{{x_ \pm }}}} \right)\frac{{{{\tilde \delta }_ + }{{\tilde \delta }_ - }}}{{\left( {1 + \tilde \delta _ + ^2} \right)\left( {1 + \tilde \delta _ - ^2} \right)}}\cos \varphi ,
\label{74}
\end{eqnarray}
\begin{eqnarray}
{D_1}\left( {{x_ \pm }} \right) = \left( {\frac{1}{{x_ \pm ^2}} + \frac{1}{{\left( {1 - {x_ \pm }} \right)^2}}} \right){b_ \pm }\left( {{x_ \pm }} \right),\: {\tilde \delta _ \pm } = 2{x_ \pm }{\delta _ \pm },
\label{75}
\end{eqnarray}
\begin{eqnarray}
{b_ \pm }\left( {{x_j}} \right) = \frac{{\tilde \delta _ \pm ^2}}{{12{{\left( {1 + \tilde \delta _ \pm ^2} \right)}^3}}}{\xi _ \pm },
\label{76}
\end{eqnarray}
\begin{eqnarray}
{\xi _ \pm } = 2\left( {1 - \tilde \delta _ \pm ^2} \right)\left( {3 - \tilde \delta _ \pm ^2} \right) - \frac{{\left( {9 + 2\tilde \delta _ \pm ^2 + \tilde \delta _ \pm ^4} \right)}}{{{x_ \pm }(1 - {x_ \pm })}} +\nonumber\\
+\left[ {\frac{{{x_ \pm }}}{{1 - {x_ \pm }}} + \frac{{1 - {x_ \pm }}}{{{x_ \pm }}}} \right]\left( {9 + 4\tilde \delta _ \pm ^2 + 3\tilde \delta _ \pm ^4} \right).
\label{77}
\end{eqnarray}
It is important to emphasize that in the differential cross sections (\ref{66})-(\ref{67}), (\ref{69}), (\ref{73}) were introduced the small corrections which are proportional to $ \sim {\left( {{m \mathord{\left/
 {\vphantom {m {{\omega _i}}}} \right.
 \kern-\nulldelimiterspace} {{\omega _i}}}} \right)^2} \ll 1$.
 These corrections make a dominant contribution to the corresponding differential cross sections under the conditions:
 \begin{eqnarray}
\left| {\varphi  - \pi } \right| \mathbin{\lower.3ex\hbox{$\buildrel<\over
{\smash{\scriptstyle\sim}\vphantom{_x}}$}} \frac{m}{{{\omega _i}}},\quad{\rm{  }}\left| {{{\tilde \delta }_ + } - {{\tilde \delta }_ - }} \right| \mathbin{\lower.3ex\hbox{$\buildrel<\over
{\smash{\scriptstyle\sim}\vphantom{_x}}$}} \frac{m}{{{\omega _i}}}.
\label{78}
\end{eqnarray}
 These conditions lead to ${D_0}\left( {{x_ \pm }} \right) \to 0,\:{\rm{ }}{d_0}\left( {{x_ \pm }} \right) \to 0$ and as a result, the corresponding differential cross sections have sharp maxima \cite{Larin}. Accordingly, the differential cross section without an external field (\ref{73}) and in the field of a wave (\ref{66}) in the kinematic region (\ref{78}) will have the following order of magnitude:
 \begin{eqnarray}
d{\sigma _{ * \left(  \pm  \right)}} \sim {Z^2}\alpha r_e^2{\left( {\frac{{{\omega _i}}}{m}} \right)^2},{\rm{  }}d{\sigma _{ \pm (r)}} \mathbin{\lower.3ex\hbox{$\buildrel<\over
{\smash{\scriptstyle\sim}\vphantom{_x}}$}} {Z^2}\alpha r_e^2{\left( {\frac{{{\omega _i}}}{m}} \right)^4}{\left( {\eta _0^r\omega \tau } \right)^2}.\quad\quad
\label{79}
\end{eqnarray}
We can see that the resonant differential cross section can be significantly larger than the corresponding cross section in the absence of a laser field.

We integrate the resonant differential cross sections (\ref{66}), (\ref{67}) and also the cross section without a laser field (\ref{73}) with respect to the azimuth angle $\varphi $:
 \begin{eqnarray}
d{\sigma _{ + (r)}} = \alpha {Z^2}r_e^2\eta _0^{2r}\frac{{r{{\left( {\omega \tau } \right)}^2}\left( {\tilde \delta _ + ^2 + \tilde \delta _ - ^2} \right){x_{ + (r)}}\left( {1 - {x_{ + (r)}}} \right)}}{{4\varepsilon _r^2G_{ + (r)}^{3/2}}}\times\nonumber\\
\times {K_{ + (r)}}P_{(r)}^{res}\left( {{\beta _ - }} \right)d{x_{ + (r)}}d\delta _ + ^2d\delta _ - ^2 ,\quad\quad
\label{80}
\end{eqnarray}
 \begin{eqnarray}
d{\sigma _{ - (r)}} = \alpha {Z^2}r_e^2\eta _0^{2r}\frac{{r{{\left( {\omega \tau } \right)}^2}\left( {\tilde \delta _ + ^2 + \tilde \delta _ - ^2} \right){x_{ - (r)}}\left( {1 - {x_{ - (r)}}} \right)}}{{4\varepsilon _r^2G_{ - (r)}^{3/2}}}\times\nonumber\\
\times{K_{ - (r)}}P_{(r)}^{res}\left( {{\beta _ + }} \right)d{x_{ - (r)}}d\delta _ + ^2d\delta _ - ^2 ,\quad\quad
\label{81}
\end{eqnarray}
where
 \begin{eqnarray}
{G_{ \pm (r)}} = {\left( {\tilde \delta _ + ^2 - \tilde \delta _ - ^2} \right)^2} + \frac{1}{2}{\left( {\frac{m}{{{\omega _i}}}} \right)^2}\left( {\tilde \delta _ + ^2 + \tilde \delta _ - ^2} \right)\times\nonumber\\
\times \left( {g_0^2\left( {{x_{ \pm (r)}}} \right) + \frac{{4{\varepsilon _r}}}{{\sin \left( {{\theta _i}/2} \right)}}\left[ {4{\varepsilon _r} - g_0^2\left( {{x_{ \pm (r)}}} \right)} \right]} \right).
\label{82}
\end{eqnarray}
Herewith, the differential cross section of PPP without a laser field takes a form\cite{Larin}:
 \begin{eqnarray}
d{\sigma _{ * \left(  \pm  \right)}} = 64{Z^2}\alpha r_e^2x_ \pm ^3{(1 - {x_ \pm })^3}\times\nonumber\\
\times \frac{{\left( {\tilde \delta _ + ^2 + \tilde \delta _ - ^2} \right){D_2}\left( {{x_ \pm }} \right)}}{{G_{ \pm \left( 0 \right)}^{3/2}}}d{x_ \pm }d\delta _ + ^2d\delta _ - ^2 ,
\label{83}
\end{eqnarray}
where ${G_{ \pm (0)}}$ can be derived from the expression for ${G_{ \pm (r)}}$ (\ref{82}) if we put ${\varepsilon _r} = 0$. Also we introduced the following expressions:
 \begin{eqnarray}
{D_2}\left( {{x_ \pm }} \right) = {D'_0}\left( {{x_ \pm }} \right) + {\left( {{m \mathord{\left/
 {\vphantom {m {{\omega _i}}}} \right.
 \kern-\nulldelimiterspace} {{\omega _i}}}} \right)^2}{D'_1}\left( {{x_ \pm }} \right),
\label{84}
\end{eqnarray}
 \begin{eqnarray}
{D'_0}\left( {{x_ \pm }} \right) =  - \frac{{\tilde \delta _ + ^2}}{{{{\left( {1 + \tilde \delta _ + ^2} \right)}^2}}} - \frac{{\tilde \delta _ - ^2}}{{{{\left( {1 + \tilde \delta _ - ^2} \right)}^2}}} +\nonumber\\
+\frac{1}{{2{x_ \pm }(1 - {x_ \pm })}}\frac{{\tilde \delta _ + ^2 + \tilde \delta _ - ^2}}{{\left( {1 + \tilde \delta _ + ^2} \right)\left( {1 + \tilde \delta _ - ^2} \right)}} +\nonumber\\
+\left( {\frac{{{x_ \pm }}}{{1 - {x_ \pm }}} + \frac{{1 - {x_ \pm }}}{{{x_ \pm }}}} \right)\frac{{2\tilde \delta _ + ^2\tilde \delta _ - ^2}}{{\left( {\tilde \delta _ + ^2 + \tilde \delta _ - ^2} \right)\left( {1 + \tilde \delta _ + ^2} \right)\left( {1 + \tilde \delta _ - ^2} \right)}},\quad
\label{85}
\end{eqnarray}
 \begin{eqnarray}
{D'_1}\left( {{x_ \pm }} \right) = {D_1}\left( {{x_ \pm }} \right) + \left( {\frac{{{x_ \pm }}}{{1 - {x_ \pm }}} + \frac{{1 - {x_ \pm }}}{{{x_ \pm }}}} \right)\times\nonumber\\
\times \frac{{g_0^2\left( {{x_ \pm }} \right)\tilde \delta _ + ^2\tilde \delta _ - ^2}}{{2\left( {\tilde \delta _ + ^2 + \tilde \delta _ - ^2} \right)\left( {1 + \tilde \delta _ + ^2} \right)\left( {1 + \tilde \delta _ - ^2} \right)}}.
\label{86}
\end{eqnarray}
The expression for the resonant profile function $P_{(r)}^{res}\left( {{\beta _ \pm }} \right)$ in (\ref{80}), when $\left| {{\beta _ \pm }} \right| \ll 1$, may be  represented as:
 \begin{eqnarray}
P_{(r)}^{res} \approx \frac{{{a_{(r)}}\Gamma _{ \pm (r)}^2}}{{\left[ {\left( {\delta _ \pm ^2 - \delta _{ \pm (r)}^2} \right) + \Gamma _{ \pm (r)}^2} \right]}} \approx P_{\max (r)}^{res} = {a_{(r)}}.
\label{87}
\end{eqnarray}
Here the parameters $\delta _{ \pm (r)}^2$ satisfy the corresponding equations for resonant frequency (for the channel A-(\ref{33}), for channel the B-(\ref{34})). The transit width for the channels A and B equals to:
 \begin{eqnarray}
{\Gamma _{ \pm (r)}} = \frac{{{c_{(r)}}{\varepsilon _r}}}{{\omega \tau }}\frac{{\left( {1 - {x_{ \pm (r)}}} \right)}}{{{x_{ \pm (r)}}}}.
\label{88}
\end{eqnarray}
We point out that the obtained expressions for the function of resonant profile (\ref{87}) are correct if transit width significantly exceeds radiation width
 \begin{eqnarray}
{\Gamma _{ \pm (r)}} \gg {\Upsilon _ \pm },
\label{89}
\end{eqnarray}
where ${\Upsilon _ \pm }$ is a radiation width:\cite{Larin}
 \begin{eqnarray}
\Upsilon _ \pm = \frac{{\alpha {\eta ^2}}}{{32\pi }}\frac{{\left( {1 - {x_ \pm }} \right)}}{{{x_ \pm }}}{K_1},
\label{90}
\end{eqnarray}
\begin{eqnarray}
{K_1} = \left( {2 + \frac{2}{{{\varepsilon _1}}} - \frac{1}{{\varepsilon _1^2}}} \right){\tanh ^{ - 1}}\left( {\sqrt {\frac{{{\varepsilon _1} - 1}}{{{\varepsilon _1}}}} } \right) -\nonumber\\
-\left( {\frac{{{\varepsilon _1} + 1}}{{{\varepsilon _1}}}} \right)\sqrt {\frac{{{\varepsilon _1} - 1}}{{{\varepsilon _1}}}}.
\label{91}
\end{eqnarray}
Taking into account the relations (\ref{88}) and (\ref{90}) the condition (\ref{89}) gives us top limitation on the duration of a laser pulse:
\begin{eqnarray}
\omega \tau  \ll {10^4}\eta _0^{ - 2}.
\label{92}
\end{eqnarray}
We underline that conditions (\ref{4}) and (\ref{92}) determine the possible intervals of the duration of a laser pulse, where the obtained differential cross sections are valid.

Let us investigate the ratio of the maximum resonant differential cross section $d\sigma _{ \pm (r)}^{\max } \approx d{\sigma _{ \pm (r)}}\left( {\left| {{\beta _ \mp }} \right| \ll 1} \right)$ (\ref{80}), (\ref{87}) to the corresponding differential cross section without a laser field (83):
\begin{eqnarray}
R_{ \pm (r)}^{\max } = \frac{{d\sigma _{ \pm (r)}^{\max }}}{{d{\sigma _{ * \left(  \pm  \right)}}}} = r{a_{(r)}}{\left( {\frac{{\eta _0^r\omega \tau }}{{64{\varepsilon _r}}}} \right)^2}\times\nonumber\\
\times \frac{{{K_{ \pm (r)}}}}{{x_{ \pm (r)}^4{{\left( {1 - {x_{ \pm (r)}}} \right)}^4}{D_2}\left( {{x_{ \pm (r)}}} \right)}}{\left[ {\frac{{{G_{ \pm (0)}}}}{{{G_{ \pm (r)}}}}} \right]^{3/2}}.
\label{93}
\end{eqnarray}

The expression (\ref{93}) determines the magnitudes of resonant differential cross section of the PPP (in the units of the corresponding differential cross section of the PPP in the absence of a laser field) for the channels A and B with simultaneous registration of the outgoing positron and electron angles (the parameters $\delta _ + ^2$ and $\delta _ - ^2$ ) and also the positron energy in the interval from ${E_{ + \left( r \right)}}$ to $\left[ {{E_{ + \left( r \right)}} + d{E_{ + \left( r \right)}}} \right]$ (for the channel A), and the electron energy in the interval from ${E_{ - \left( r \right)}}$ to $\left[ {{E_{ - \left( r \right)}} + d{E_{ - \left( r \right)}}} \right]$ (for the channel B). It is important to emphasize that for the channel A the positron outgoing angle relative to the initial gamma quantum momentum (parameter $\delta _ + ^2$) determines as the positron resonant energy ${E_{ + \left( r \right)}}$ , so the electron resonant energy ${E_{ - \left( r \right)}} \approx {\omega _i} - {E_{ + \left( r \right)}}$. Herewith, these values do not depend on the electron outgoing angle (parameter$\delta _ - ^2$). Meanwhile, for the channel B we have opposite situation. Herein the electron outgoing angle relative to the initial gamma quantum momentum (parameter $\delta _ - ^2$)  determines as the electron resonant energy ${E_{ - \left( r \right)}}$, so the positron resonant energy ${E_{ + \left( r \right)}} \approx {\omega _i} - {E_{ - \left( r \right)}}$ and these values do not depend on the parameter $\delta _ + ^2$.
\begin{figure}[ht]
\includegraphics[width=1.1\linewidth]{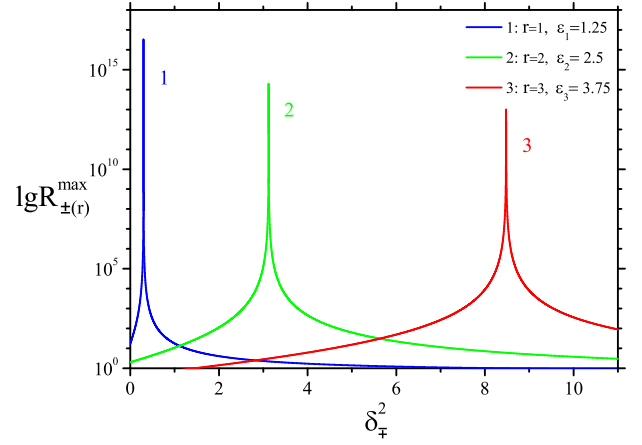}%
\caption{Dependence of the function $R_{ + (r)}^{\max }$ (\ref{93}) on the parameter $\delta _ - ^2$ at the fixed value of the parameter $\delta _ + ^2 = 0.1$ (for the channel A) and dependence of the function $R_{ - (r)}^{\max }$ on the parameter $\delta _ + ^2$ at the fixed value of the parameter $\delta _ - ^2 = 0.1$ (for the channel B) for the first three resonances and the initial gamma quantum energy ${\omega _i} = 125\;{\rm{GeV }}\left( {{\omega _i} > {\omega _{thr\left( 1 \right)}} = 100\;{\rm{GeV}}} \right)$. \label{Fig7}}%
\end{figure}

In Fig.(\ref{Fig7}) we represent dependences of the functions $R_{ \pm (r)}^{\max }$ (\ref{93}) on the electron or positron outgoing angles (parameters $\delta _ \mp ^2$) with the certain value of the initial gamma quantum energy and the fixed outgoing angle of a positron (for the channel A) or an electron (for the channel B). We can conclude that when $\tilde \delta _ - ^2 = \tilde \delta _ + ^2$ (\ref{72}) (for the channel A) and  $\tilde \delta _ + ^2 = \tilde \delta _ - ^2$ (for the channel B) there is a sharp maximum in the resonant differential cross section. The resonant differential cross section significantly exceeds the corresponding one of the PPP without a laser field. For the first resonance $R_{ \pm (1)}^{\max } \mathbin{\lower.3ex\hbox{$\buildrel>\over
{\smash{\scriptstyle\sim}\vphantom{_x}}$}} {10^{16}}$, for the second resonance $R_{ \pm (2)}^{\max } \mathbin{\lower.3ex\hbox{$\buildrel>\over
{\smash{\scriptstyle\sim}\vphantom{_x}}$}} {10^{14}}$ and for the third resonance $R_{ \pm (3)}^{\max } \mathbin{\lower.3ex\hbox{$\buildrel>\over
{\smash{\scriptstyle\sim}\vphantom{_x}}$}} {10^{12}}$ (see (\ref{79})). We accentuate that in Fig.(\ref{Fig7}) all curves are depicted for the sign "+" in front of the square root in the expression for the positron energy (\ref{33}) (the channel A) or the electron energy (\ref{34}) (the channel B). The second solution with sign "-"  suppressed and does not make a significant contribution to the value of the resonant differential cross section.

We integrate the resonant differential cross section with respect to the parameter $\delta _ - ^2$ (for the channel A) and parameter $\delta _ + ^2$ (for the channel B):
\begin{eqnarray}
d{\sigma '_{ + (r)}} = \left( {\alpha r_e^2{Z^2}} \right){g_r}{F_{ + (r)}}d{x_{ + (r)}}d\delta _ + ^2,
\label{94}
\end{eqnarray}
\begin{eqnarray}
d{\sigma '_{ - (r)}} = \left( {\alpha r_e^2{Z^2}} \right){g_r}{F_{ - (r)}}d{x_{ - (r)}}d\delta _ - ^2,
\label{95}
\end{eqnarray}
where
\begin{eqnarray}
{g_r} = {\left( {\frac{{{\omega _{thr\left( r \right)}}}}{m}} \right)^2}{\left( {\eta _0^r\omega \tau } \right)^2},
\label{96}
\end{eqnarray}
\begin{eqnarray}
{F_{ + (r)}} = \frac{r}{{{\varepsilon _r}}}\left( {\frac{{{x_{ + (r)}}}}{{1 - {x_{ + (r)}}}}} \right){K_{ + (r)}}P_{(r)}^{res}\left( {{\beta _ - }} \right),
\label{97}
\end{eqnarray}
\begin{eqnarray}
{F_{ - (r)}} = \frac{r}{{{\varepsilon _r}}}\left( {\frac{{{x_{ - (r)}}}}{{1 - {x_{ - (r)}}}}} \right){K_{ - (r)}} P_{(r)}^{res}\left( {{\beta _ + }} \right).
\label{98}
\end{eqnarray}
The relation (\ref{94}) defines the resonant differential cross section of the PPP with simultaneous registration of the positron outgoing angle and the positron energy (irrespective to the electron outgoing angles). The relation (\ref{95}) defines the resonant differential cross section of the PPP with simultaneous registration of the outgoing electron angle and the electron energy (irrespective to the positron outgoing angles). We underline that the function ${g_r}$ (96) is the same for both channels. It is defined by two factors. One is large enough and associated with the small transferred momentum and the threshold energy of $r^{th}$-resonance . Other is associated with the transit width.  For example, when ${\omega _{thr\left( 1 \right)}} = 100\;{\rm{GeV}}$, ${\eta _0} = {10^{ - 1}}$, $\left( {\omega \tau } \right) = {10^2}$ from (\ref{96}) we can calculate that: for the first resonance ${g_1} \approx 4 \cdot {10^{12}},$ for the second resonance ${g_2} \approx {10^{10}},$ for the third resonance ${g_3} \approx 4.4 \cdot {10^7}$, for the fourth resonance ${g_4} \approx 2.5 \cdot {10^5}$, for the fifth resonance ${g_5} \approx 1.6 \cdot {10^3}$. Here function ${g_r}$ essentially depends on the wave intensity and also on the resonance number. There is suppression of the laser-stimulated Breit-Wheeler process with the increase in the number of absorbed photons and the decrease in the wave intensity. It is important to point out that for the certain resonance (certain $r$) the quantitative difference of the resonant differential cross sections for the channels A and B is determined by the functions ${F_{ \pm (r)}}$ (\ref{97}), (\ref{98}). When $\left| {{\beta _ - }} \right| \ll 1,{\rm{ }}\left| {{\beta _ + }} \right| \ll 1$ these functions and the corresponding resonant differential cross sections (\ref{94}), (\ref{95}) are of the maximum values:
\begin{eqnarray}
d\sigma _{ + (r)}^{\max } = \left( {\alpha r_e^2{Z^2}} \right)\Phi _{ + (r)}^{\max }d{x_{ + (r)}}d\delta _ + ^2,
\label{99}
\end{eqnarray}
\begin{eqnarray}
d\sigma _{ - (r)}^{\max } = \left( {\alpha r_e^2{Z^2}} \right)\Phi _{ - (r)}^{\max }d{x_{ - (r)}}d\delta _ - ^2,
\label{100}
\end{eqnarray}
\begin{eqnarray}
\Phi _{ \pm (r)}^{\max } = {g_r}F_{ \pm (r)}^{\max },
\label{101}
\end{eqnarray}
\begin{eqnarray}
F_{ \pm (r)}^{\max } = \frac{{r{a_{(r)}}}}{{\varepsilon _r^2}}\left( {\frac{{{x_{ \pm (r)}}}}{{1 - {x_{ \pm (r)}}}}} \right){K_{ \pm (r)}}.
\label{102}
\end{eqnarray}

\begin{figure}
\includegraphics[width=1.1\linewidth]{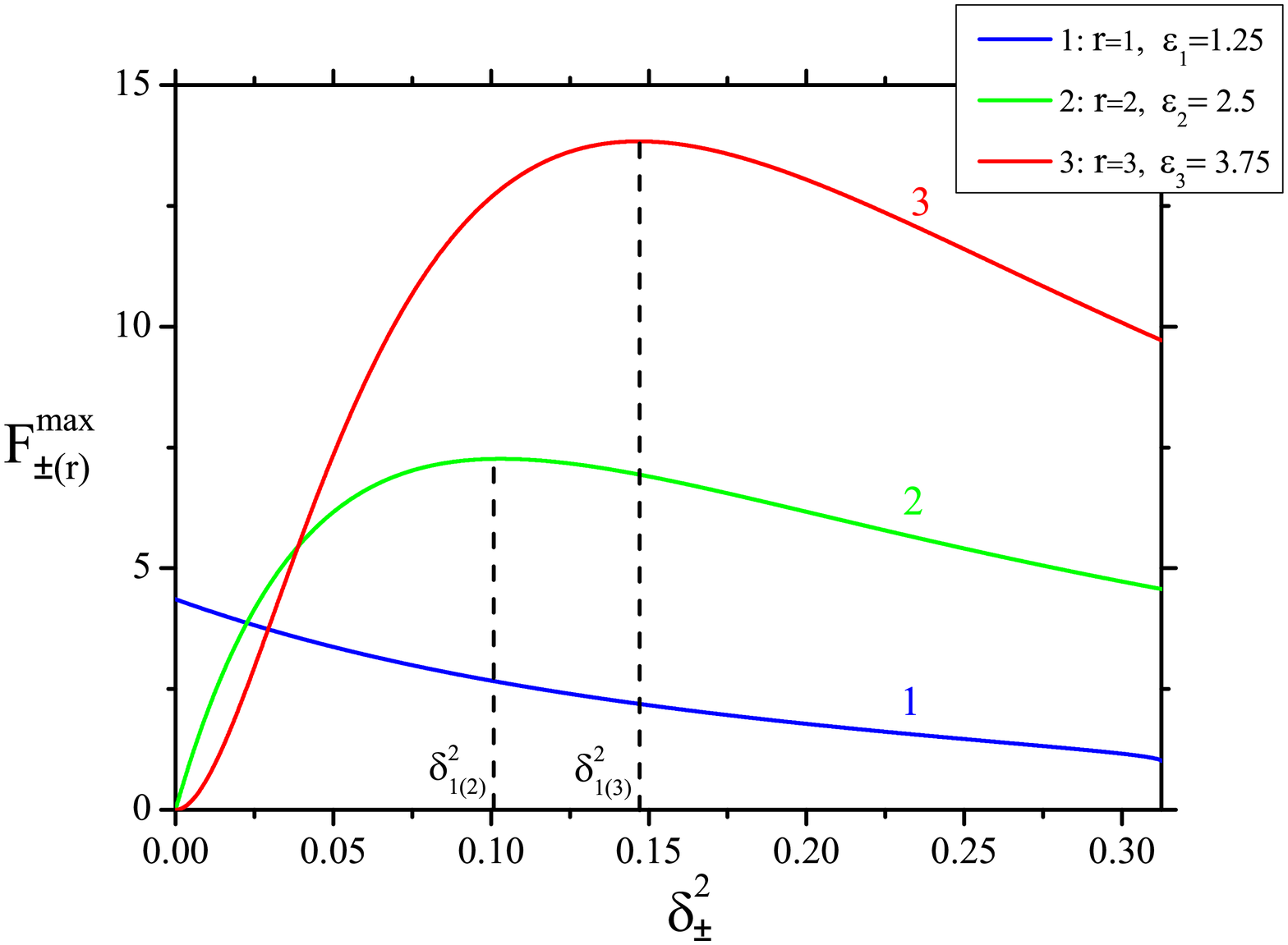}%
\caption{Dependences of the functions $F_{ \pm (r)}^{\max }$ (\ref{102}) on the parameters $\delta _ \pm ^2$  for the first three resonances with the certain energy of the initial gamma quantum ${\omega _i} = 125\;{\rm{GeV  }}\left( {{\omega _i} > {\omega _{thr\left( 1 \right)}} = 100\;{\rm{GeV}}} \right)$.\label{Fig8}}%
\end{figure}
\begin{figure}
\includegraphics[width=1.1\linewidth]{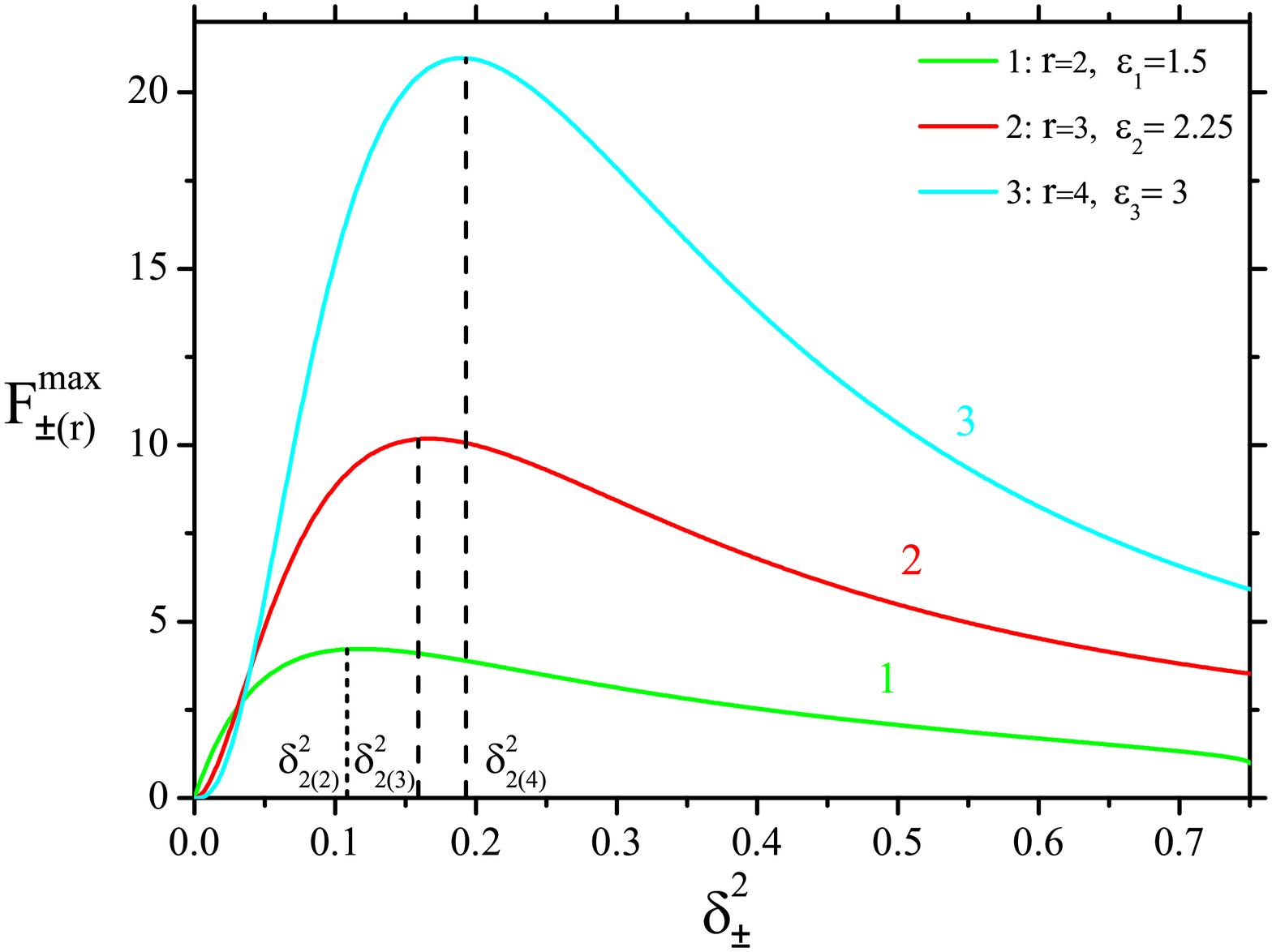}%
\caption{Dependences of the functions $F_{ \pm (r)}^{\max }$ (\ref{102}) on the parameters $\delta _ \pm ^2$  for the first three resonances with the certain energy of the initial gamma quantum ${\omega _i} = 75\;{\rm{GeV}}$ $\left( {{\omega _{thr\left( 1 \right)}} > {\omega _i} > {\omega _{thr\left( 2 \right)}},{\rm{  }}\:{\omega _{thr\left( 2 \right)}} = 50\;{\rm{GeV}}} \right).$ \label{Fig9}}%
\end{figure}
\begin{figure}
\includegraphics[width=1.1\linewidth]{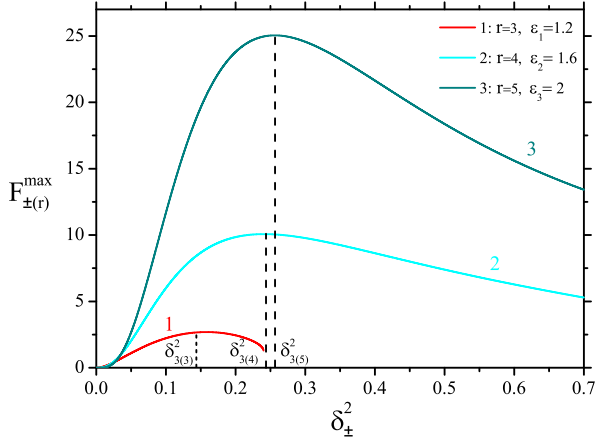}%
\caption{Dependences of the functions $F_{ \pm (r)}^{\max }$ (\ref{102}) on the parameters $\delta _ \pm ^2$  for the first three resonances with the certain energy of the initial gamma quantum ${\omega _i} = 40\;{\rm{GeV}}$ $\left( {{\omega _{thr\left( 2 \right)}} > {\omega _i} > {\omega _{thr\left( 3 \right)}},\quad {\omega _{thr\left( 3 \right)}} = 33.3\;{\rm{GeV}}} \right)$. \label{Fig10}}%
\end{figure}

In Fig.(\ref{Fig8}), Fig.(\ref{Fig9}), Fig.(\ref{Fig10}) we represent the dependences of the functions $F_{ \pm (r)}^{\max }$ (102) on the corresponding parameters $\delta _ \pm ^2$ (the positron and the electron outgoing angle) for the first three possible resonances for the different initial gamma quantum energies. We can see that the all resonances, except first, $\left( {r = 2,3,4,5, \ldots } \right)$ have maxima in the distribution functions. These maxima correspond to the most probable values of the electron-positron pair energies for the channels A and B at the corresponding outgoing angles $\delta _ + ^2 = \delta _{ j \left( r \right)}^2$ or $\delta _ - ^2 = \delta _{ j \left( r \right)}^2$, where $j$ stands for the different gamma quantum energies ($j=1$ for ${\omega _i} = 125{\rm{ GeV}}$, $j=2$ for ${\omega _i} = 75{\rm{ GeV}}$, $j=3$ for ${\omega _i} = 40{\rm{ GeV}}$). The represented figures correspond to the positron (electron) energy with the sign "+" in front of the square root (\ref{33}), (\ref{34}). The second solution for the positron (electron) energy (with sign "-") does not make an essential contribution to these functions.

\begin{table*}
\caption{The most probable values of the electron-positron pair energies and the corresponding resonant differential cross sections.\label{Tab1}}
\begin{tabular}{@{}|c|c|c|c|c|c|@{}}
\toprule
\multicolumn{6}{|c|}{${\omega _i} = 125{\rm{ GeV,}}\quad {\omega _{thr\left( 1 \right)}} = 100{\rm{ GeV}}$}                                         \\ \midrule
r                  & channel & $\delta _{ \pm (r)}^2$ & ${E_ + },{\rm{ GeV}}$ & ${E_ - },{\rm{ GeV}}$ & $\Phi _{ \pm (r)}^{\max }{\rm{ }}\left( {\eta  = 0.1} \right)$                  \\ \midrule
\multirow{2}{*}{1} & A       & $0 \le \delta _ + ^2 \le 0.3125$     & $34.55 \le {E_ + } \le 90.45$  & ${E_ - } = \left( {125 - {E_ + }} \right)$  & \multirow{2}{*}{${\rm{4}} \cdot {\rm{1}}{{\rm{0}}^{12}} \le \Phi _{ \pm (1)}^{\max } \le {\rm{1}}{\rm{.7}} \cdot {\rm{1}}{{\rm{0}}^{13}}$} \\ \cmidrule(lr){2-5}
                   & B       & $0 \le \delta _ - ^2 \le 0.3125$     & ${E_ + } = \left( {125 - {E_ - }} \right)$  & $34.55 \le {E_ - } \le 90.45$  &                    \\ \midrule
\multirow{2}{*}{2} & A       & $\delta _{ + (2)}^2 = \delta _{1(2)}^2 = 0.103$     & $105.88$  & $19.12$  & \multirow{2}{*}{$\Phi _{ \pm (2)}^{\max } \approx 6.95 \cdot {10^{10}}$} \\ \cmidrule(lr){2-5}
                   & B       & $\delta _{ - (2)}^2 = \delta _{1(2)}^2 = 0.103$     & $19.12$  & $105.88$  &                    \\ \midrule
\multirow{2}{*}{3} & A       & $\delta _{ + (3)}^2 = \delta _{1(3)}^2 = 0.147$     & $111.28$  & $13.72$  & \multirow{2}{*}{$\Phi _{ \pm (3)}^{\max } \approx 5.89 \cdot {10^8}$} \\ \cmidrule(lr){2-5}
                   & B       & $\delta _{ - (3)}^2 = \delta _{1(3)}^2 = 0.147$     & $13.72$  & $111.28$  &                    \\ \bottomrule
\end{tabular}
\end{table*}
\begin{table}
\caption{The most probable values of the electron-positron pair energies and the corresponding resonant differential cross sections.\label{Tab2}}
\begin{tabular}{@{}|c|c|c|c|c|c|@{}}
\toprule
\multicolumn{6}{|c|}{${\omega _i} = 75{\rm{ GeV,}}\quad {\omega _{thr\left( 2 \right)}} = 50{\rm{ GeV}}$}                                         \\ \midrule
r                  & channel & $\delta _{ \pm (r)}^2$ & ${E_ + },{\rm{ GeV}}$ & ${E_ - },{\rm{ GeV}}$ & ${\rm{   }}\Phi _{ \pm (r)}^{\max }{\rm{ }}\left( {\eta  = 0.1} \right)$                  \\ \midrule
\multirow{2}{*}{1} & A       & $\delta _ + ^2 = \delta _{2(2)}^2 = 0.118$     & $53.19$  & $21.81$  & \multirow{2}{*}{$\Phi _{ \pm (2)}^{\max } \approx 4.05 \cdot {10^{10}}$} \\ \cmidrule(lr){2-5}
                   & B       & $\delta _ - ^2 = \delta _{2(2)}^2 = 0.118$     & $21.81$  & $53.19$  &                    \\ \midrule
\multirow{2}{*}{2} & A       & $\delta _ + ^2 = \delta _{2(3)}^2 = 0.167$     & $60.15$  & $14.85$  & \multirow{2}{*}{$\Phi _{ \pm (3)}^{\max } \approx 4.33 \cdot {10^8}$} \\ \cmidrule(lr){2-5}
                   & B       & $\delta _ - ^2 = \delta _{2(3)}^2 = 0.167$     & $14.85$  & $60.15$  &                    \\ \midrule
\multirow{2}{*}{3} & A       & $\delta _ + ^2 = \delta _{2(4)}^2 = 0.19$     & $63.6$  & $11.4$  & \multirow{2}{*}{$\Phi _{ \pm (4)}^{\max } \approx 5.02 \cdot {10^6}$} \\ \cmidrule(lr){2-5}
                   & B       & $\delta _ - ^2 = \delta _{2(4)}^2 = 0.19$     & $11.4$  & $63.6$  &                    \\ \bottomrule
\end{tabular}
\end{table}
\begin{table}
\caption{The most probable values of the electron-positron pair energies and the corresponding resonant differential cross sections.\label{Tab3}}
\begin{tabular}{@{}|c|c|c|c|c|c|@{}}
\toprule
\multicolumn{6}{|c|}{${\omega _i} = 40{\rm{ GeV,}}\quad {\omega _{thr\left( 3 \right)}} = 33.3{\rm{ GeV}}$}                                         \\ \midrule
r                  & channel & $\delta _{ \pm (r)}^2$ & ${E_ + },{\rm{ GeV}}$ & ${E_ - },{\rm{ GeV}}$ & ${\rm{   }}\Phi _{ \pm (r)}^{\max }{\rm{ }}\left( {\eta  = 0.1} \right)$                  \\ \midrule
\multirow{2}{*}{1} & A       & $\delta _ + ^2 = \delta _{3(3)}^2 = 0.145$     & $22.41$  & $17.59$  & \multirow{2}{*}{$\Phi _{ \pm (3)}^{\max } \approx 9.29 \cdot {10^7}$} \\ \cmidrule(lr){2-5}
                   & B       & $\delta _ - ^2 = \delta _{3(3)}^2 = 0.145$     & $17.59$  & $22.41$  &                    \\ \midrule
\multirow{2}{*}{2} & A       & $\delta _ + ^2 = \delta _{3(4)}^2 = 0.241$     & $26.6$  & $13.4$  & \multirow{2}{*}{$\Phi _{ \pm (4)}^{\max } \approx 2.41 \cdot {10^6}$} \\ \cmidrule(lr){2-5}
                   & B       & $\delta _ - ^2 = \delta _{3(4)}^2 = 0.241$     & $13.4$  & $26.6$  &                    \\ \midrule
\multirow{2}{*}{3} & A       & $\delta _ + ^2 = \delta _{3(5)}^2 = 0.256$     & $29.43$  & $10.57$  & \multirow{2}{*}{$\Phi _{ \pm (5)}^{\max } \approx 3.84 \cdot {10^4}$} \\ \cmidrule(lr){2-5}
                   & B       & $\delta _ - ^2 = \delta _{3(5)}^2 = 0.256$     & $10.57$  & $29.43$  &                    \\ \bottomrule
\end{tabular}
\end{table}

In the Tab.\ref{Tab1}-Tab.\ref{Tab3} for the channels A and B for the first several resonances we represent the most probable values of the electron-positron pair energy at the corresponding outgoing angles of a positron or an electron $\left( {\delta _{ j \left( r \right)}^2} \right)$ and the magnitude of the resonant differential cross section (in the units of $\alpha r_e^2{Z^2}$) $\Phi _{ \pm (r)}^{\max }$ for the different initial gamma quantum energies. We can see from these tables that for the all initial gamma quanta energies $\left( {{\omega _i} = 125{\rm{ GeV,}}\;75{\rm{ GeV}},\;40{\rm{ GeV}}} \right)$ the most probable energies of the positron and the electron in the frame of same resonance are differ.  When we transfer from the channel A to channel B these energies replace each other $\left( {{E_{ + \left( A \right)}} \to {E_{ - \left( B \right)}},\;{E_{ - \left( A \right)}} \to {E_{ + \left( B \right)}}} \right)$. Moreover, for the channel A the positron energy is greater than the electron energy. However, for the channel B the electron energy is greater than the positron energy. This fact gives us the opportunity to determine the appropriate channel in an experiment with measurement of the electron-positron pair energies. It is important to note that there are different sets of the electron-positron pair energies for the different channels. In addition, the corresponding differential cross sections are large enough (see value of $\Phi _{ \pm (r)}^{\max }$ in Tab.\ref{Tab1}-Tab.\ref{Tab3}). The order of magnitudes for the obtained resonant differential cross sections (in the units of $\alpha r_e^2{Z^2}$) varies in the wide range: from  $ \sim {10^{13}}$ to $ \sim {10^{8}}$ (for ${\omega _i} = 125{\rm{ GeV}}$). The magnitude of the resonant differential cross section decreases with the decrease in the initial gamma quantum energy. Nevertheless, it still remains large enough. For example for the initial gamma quantum energy ${\omega _i} = 40{\rm{ GeV}}$ we have $\Phi _{ \pm (3)}^{\max } \sim {10^8}$ (for the third resonance) and $\Phi _{ \pm (5)}^{\max } \sim {10^4}$ (for the fifth resonance).

\section{Conclusion}
The study of the resonant process of the PPP by a high-energy gamma quantum in the field of a nucleus and a weak quasi-monochromatic laser wave allows us to formulate the main results.

The process of the resonant PPP for the channels A and B effectively reduces into the two processes of the first order in the fine structure constant: the production process of the ultrarelativistic electron-positron pair by the gamma quantum due to absorption of $r$-photons of a wave (the laser-stimulated Breit-Wheeler process) and the process of the intermediate electron (positron) scattering on a nucleus in the wave field (the laser-assisted Mott scattering). 

There is the threshold energy of the process which depends on the number $r$ of absorbed photons of a laser wave ${\omega _{thr\left( r \right)}}$ (\ref{31}), where $r = 1,2,3, \ldots $ is the resonance number. For the first resonance within the optical frequency range, this value is of the order of ${\omega _{thr\left( 1 \right)}} \sim {10^2}\;{\rm{GeV}}$. The threshold energy decreases with the increase in the resonance number ${\omega _{thr\left( r \right)}} = {{{\omega _{thr\left( 1 \right)}}} \mathord{\left/
 {\vphantom {{{\omega _{thr\left( 1 \right)}}} r}} \right.
 \kern-\nulldelimiterspace} r}$. We studied the resonant photoproduction of the ultrarelativistic electron-positron pair, when all particles (the initial gamma quantum and the electron-positron pair) propagate in a narrow cone at a large angle relative to direction of a wave propagation.
 
The outgoing angle of a positron (for the channel A) or an electron (for the channel B) relative to the initial gamma quantum momentum defines the electron-positron pair energy, which significantly depends on the number of absorbed wave photons (the resonance number). Herewith, for the channel A the most likely the positron energy is greater than the electron one. For the channel B we have opposite situation, the most probable electron energy is greater than the positron one. This fact allows us to distinguish the corresponding reaction channel by the electron-positron pair energy.
 
The distributions of the resonant differential cross sections over the outgoing angles of a positron or an electron of the higher resonances ($r = 2,3, \ldots $) have clearly defined maxima, in contrast to the corresponding distribution of the first resonance. The most probable positron and electron energies significantly differ from each other as for the frame of the same resonance, so for the different resonances.

For the intensities of a laser wave $I \mathbin{\lower.3ex\hbox{$\buildrel<\over
{\smash{\scriptstyle\sim}\vphantom{_x}}$}} {10^{16}} \div {10^{17}}{{\rm{W}} \mathord{\left/
 {\vphantom {{\rm{W}} {{\rm{c}}{{\rm{m}}^{\rm{2}}}}}} \right.
 \kern-\nulldelimiterspace} {{\rm{c}}{{\rm{m}}^{\rm{2}}}}}$ and the initial gamma quantum energy ${\omega _i} = 125\;{\rm{GeV}}$ the resonant energies of an electron-positron pair for the channels A and B in the case of first, second and third resonances may be measured with an extremely large magnitude of the differential cross section: from $ \sim {10^{13}}$ for the first resonance to $ \sim {10^8}$ (in the units of $\alpha {Z^2}r_e^2$) for the third resonance (see Tab.\ref{Tab1}).
 
The obtained results may be experimentally verified using the facilities of pulsed laser radiation (SLAC, FAIR, XFEL, ELI, XCELS).

\bibliography{APR}

\end{document}